J. Mech. Phys. Solids 195 (2025) 105955

Contents lists available at ScienceDirect

# Journal of the Mechanics and Physics of Solids

journal homepage: www.elsevier.com/locate/jmps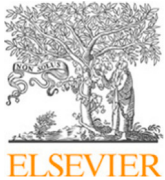
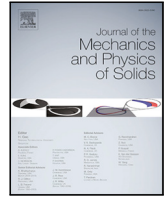

# Finite strain continuum phenomenological model describing the shape-memory effects in multi-phase semi-crystalline networks

Matteo Arricca [a,*], Nicoletta Inverardi [b], Stefano Pandini [b], Maurizio Toselli [c], Massimo Messori [d], Giulia Scalet [a,*]

[a] *Department of Civil Engineering and Architecture, University of Pavia, via Ferrata 3, Pavia 27100, Italy*
[b] *Department of Mechanical and Industrial Engineering, University of Brescia, via Branze 38, Brescia 25133, Italy*
[c] *Department of Industrial Chemistry "Toso Montanari", University of Bologna, Viale Risorgimento 4, Bologna 40136, Italy*
[d] *Department of Applied Science and Technology, Politecnico di Torino, Corso Duca degli Abruzzi 24, Torino 10129, Italy*## A R T I C L E  I N F O

Dataset link: https://doi.org/10.5281/zenodo.14196360

Keywords:
Shape-memory polymers
Semi-crystalline polymer networks
Two-way shape-memory effect
Phenomenological modeling
Constitutive formulation
Finite strain mechanics## A B S T R A C T

Thermally-driven semi-crystalline polymer networks are capable to achieve both the one-way shape-memory effect and two-way shape-memory effect under stress and stress-free conditions, therefore representing an appealing class of polymers for applications requiring autonomous reversible actuation and shape changes. In these materials, the shape-memory effects are achieved by leveraging the synergistic interaction between one or more crystalline phases and the surrounding amorphous ones that are present within the network itself. The present paper introduces a general framework for the finite strain continuum phenomenological modeling of the thermo-mechanical and shape-memory behavior of multi-phase semi-crystalline polymer networks. Model formulation, including the definition of phase and control variables, kinematic assumptions, and constitutive specifications, is introduced and thoroughly discussed. Theoretical derivations are general and easily adaptable to all cross-linked systems which include two or more crystalline domains or a single crystalline phase with a wide melting range and manifest macroscopically the one-way shape-memory effect and the two-way shape-memory effect under stress and stress-free conditions. Model capabilities are validated against experimental data for copolymer networks with two different crystalline phases characterized by well-separated melting and crystallization transitions. Results demonstrate the accuracy of the proposed model in predicting all the phenomena involved and in furnishing a useful support for future material and application design purposes.## 1. Introduction

Shape-memory polymers (SMPs) are smart materials characterized by the unique property of recovering their original (permanent) shape from a temporarily deformed one under exposure to an external stimulus such as heat, light, or a magnetic field (Hager et al., 2015). In view of their advantages, such as light weight, low cost, good processability, potential biocompatibility, high shape deformability and recoverability, SMPs have found applications in several fields, *e.g.*, intelligent medical devices, sensors and actuators, self-deployable structures in spacecraft (see Meng and Hu (2009) and references therein), and 4D fabrication when combined with additive manufacturing techniques (Jiang et al., 2022; Bonetti et al., 2024).

* Corresponding authors.
*E-mail addresses:* matteo.arricca@unipv.it (M. Arricca), giulia.scalet@unipv.it (G. Scalet).https://doi.org/10.1016/j.jmps.2024.105955
Received 25 September 2024; Received in revised form 11 November 2024; Accepted 11 November 2024
Available online 19 November 20240022-5096/© 2024 The Authors. Published by Elsevier Ltd. This is an open access article under the CC BY license (http://creativecommons.org/licenses/by/4.0/).

The capability of recovering the permanent shape, provided via material processing, from a previously imposed temporary shape, is termed shape-memory effect (SME). Particularly, the one-way SME represents a non-reversible feature in view of the necessity of resorting to an external mechanical intervention to set again the temporary shape from the (recovered) permanent one (Scalet, 2020; Arricca et al., 2024). Rather, the two-way SME ensures reversibility between two distinguished temporary shapes upon the application of an on-off stimulus, as a cooling–heating cycle for instance, therefore displaying the capability of a reversible shape effect between two different configurations. Such an effect becomes highly appealing for applications requiring autonomous reversible actuation and shape changes, *e.g.*, artificial muscles (Fan and Li, 2017), reconfigurable structures (Risso et al., 2024), and soft robots (Scalet, 2020), and it is therefore object of both scientific and commercial interests.

There exist various strategies that may be conceived to induce the thermally-driven two-way SME under stress and/or stress-free conditions (Wang et al., 2019). Multi-phase semi-crystalline networks represent an attractive material, given their easy tailorability, good shape-memory performances, and fast responses to triggering stimuli (Lendlein and Langer, 2002; Scalet et al., 2018).

The term "multi-phase semi-crystalline networks" includes all cross-linked systems with two or more crystalline domains or a single crystalline phase with a wide melting range (*i.e.*, systems composed by a broad continuous series of different crystalline domains). Thanks to the presence of chemical cross-links and of at least one crystallizable phase, these networks are capable to macroscopically manifest the one-way SME and two-way SME under stress and stress-free conditions. Particularly, these SMEs can be achieved by leveraging the synergistic interaction between the crystalline and amorphous phases present within the network itself. In fact, in these systems the traditional two-way SME is commonly obtained by means of cooling–heating cycles from above the melting temperature(s) to below the crystallization temperature(s) under an external applied stress (Chung et al., 2008; Resnina et al., 2015; Basak and Bandyopadhyay, 2022; Feng and Li, 2022). In such a case, reversible actuation is associated to an elongation of the material across the crystallization temperature(s) during cooling (known as crystallization-induced elongation (CIE)) and to a contraction across the melting temperature(s) during heating (known as melting-induced contraction (MIC)) (Chung et al., 2008; Scalet et al., 2018). The melting transition(s) is thus employed to trigger shape recovery, while temporary shapes are fixed by crystallization(s).

Additionally, a two-way SME under stress-free conditions can also be achieved in these systems and it arises from a generated internal stress. In fact, its achievement is allowed by the presence of two phases that work synergistically: a skeleton phase, which provides the necessary internal stress, and an actuation phase, which is responsible for the actuation by undergoing a reversible shape change upon temperature-induced crystallization-melting cycles under the generated internal stress (Arricca et al., 2024). In the specific case of semi-crystalline networks, the phase(s) melted during heating acts as actuation phase, while the unmelted (*i.e.*, crystalline) phase(s) is the skeleton phase. The response strongly depends on the relative contents of these two phases.

Accordingly, various multi-phase semi-crystalline networks have been proposed, such as semi-crystalline homopolymer networks, *i.e.*, having one crystalline and one amorphous phase (Behl et al., 2013; Turner et al., 2014; Zhou et al., 2014; Tippets et al., 2015; Dolynchuk et al., 2017; Wang et al., 2017; Xu et al., 2019; Wang et al., 2020; Yuan et al., 2020; Hao et al., 2022; Huang et al., 2023; Jiang et al., 2022; Wang et al., 2022; Xu et al., 2022; Ren and Feng, 2023; Wong et al., 2023), semi-crystalline copolymer networks, *i.e.*, having two crystalline and two amorphous phases (Garle et al., 2012; Saatchi et al., 2015; Wang et al., 2017; Fan et al., 2018; Yan et al., 2018; Ilić-Stojanović et al., 2021; Inverardi et al., 2022; Li et al., 2022; Xiang et al., 2022; He et al., 2024; Xu et al., 2024), and multi-phase semi-interpenetrating networks, *i.e.*, having more than one crystalline and amorphous phases (Uto et al., 2023). As an alternative to complex interpenetrating or block-copolymer networks, Kolesov et al. (2015) investigated covalently cross-linked binary and ternary blends having, respectively, two and three crystalline phases with different thermal stability.

Although the synthesis of these material systems has been object of exhaustive studies, the modeling of all their SMEs to support the design of applications, and the understanding of the complex phenomena behind, still needs to be investigated. To the best of authors' knowledge, no modeling approaches are available for the two-way shape-memory behavior of multi-phase semi-crystalline networks. In fact, most of the models were proposed using a phase transition approach to describe the shape-memory behavior of cross-linked networks with a single crystalline phase in a one-dimensional and three-dimensional framework (Westbrook et al., 2010; Dolynchuk et al., 2014, 2015; Scalet et al., 2018; Prasad et al., 2021; Yan et al., 2020; Zeng et al., 2021, 2023; Arricca et al., 2024; Gülaşık et al., 2024). Recently, Gu et al. (2024) proposed a thermodynamic model with internal state variables to describe only the two-way SME under an applied stress of semi-crystalline networks with two crystalline domains.

The present paper aims to propose a general finite strain continuum formulation for the phenomenological modeling of the SMEs in multi-phase semi-crystalline networks under stress and stress-free conditions.

We adopt a phase-transition approach that has been applied to a wide variety of SMPs, also coupled with physically-based phase evolution laws (Liu et al., 2006; Guo et al., 2016; Yang and Li, 2016; Arricca et al., 2024), thanks to its effectiveness. Accordingly, the architecture of the SMP network is represented at the macroscopic scale by continuum variables to describe the amorphous and crystalline phases and the inelastic contributions associated with the shape-memory deformation. Constitutive equations are stated to establish the mechanical stress–strain response of the network, and evolution laws are derived on the basis of physical evidences arising from previous works (Scalet et al., 2018; Arricca et al., 2024) and experimental observations (Inverardi et al., 2022). Theoretical derivations are general and easily adaptable to all cross-linked systems which include two or more crystalline domains or a single crystalline phase with a wide melting range and manifest macroscopically the one-way SME and two-way SME under stress and stress-free conditions. Model capabilities are validated against uniaxial experimental data performed by Inverardi et al. (2022) on poly(ethylene glycol)/poly($\varepsilon$-caprolactone) copolymer networks with two different crystalline phases characterized by well-separated melting and crystallization transitions. Furthermore, we perform additional testing compared to Inverardi et al. (2022) to have a comprehensive set of experimental data for the identification of model parameters. Results demonstrate the accuracy of the proposed model in predicting all the phenomena involved and in furnishing a useful support for future material and application design purposes.

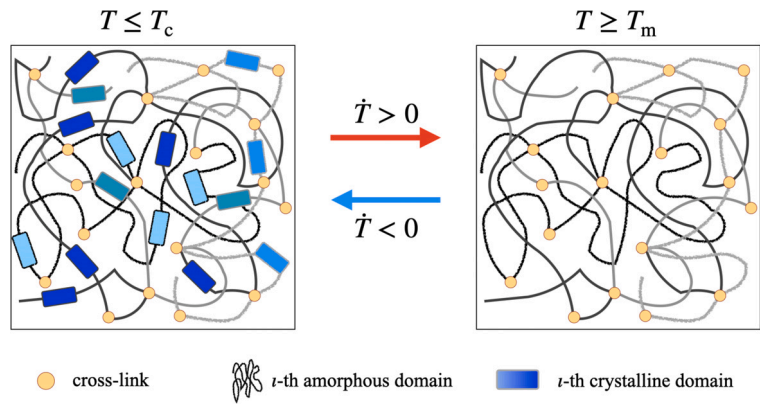

**Fig. 1.** Schematic of the internal architecture evolution of a multi-phase semi-crystalline network, occurring via cooling-heating cycles. For the sake of generality, the amorphous domains (random coiled lines) and the crystalline contents (solid shapes) are represented with $n$ different shades of gray and blue colors, respectively, to distinguish between the $n$ polymers constituting the network. Orange circles depict the cross-links.

The manuscript is organized as follows. Section 2 introduces the material system and identifies the necessary continuum variables to describe the phase contents. Section 3 introduces kinematic specifications in terms of multiplicative decomposition of the deformation gradient and characterization of the mechanical deformation in the amorphous and crystalline contents. The constitutive stress–strain response of the system is introduced in Section 4, with further useful insights and clarifications on the selection of the energy functions. Section 5 presents the application of the theoretical formulation to a copolymer network, together with the validation of the derived equations on experimental data. Concluding remarks in Section 6 complete the paper.

## 2. Material and continuum variables

This section presents the shape-memory material system under investigation and defines the continuum variables to describe the phase contents.

### 2.1. Material system description

Consider a multi-phase semi-crystalline network made of $n$ semi-crystalline polymers, being $n \geq 1$ a positive integer. For the sake of generality, in the following we do not specify the number $n$ of polymers. Each $\iota$-th polymer ($\iota = 1, \ldots, n$) is generally described as composed of one amorphous domain, and (potentially) one crystalline domain as a function of the temperature, as clarified hereafter. We also refer to the amorphous and crystalline domains as phases. Chemical cross-links are established among the polymer chains within the amorphous region and act as net-points responsible for determining the permanent shape of the entire network during shape-memory tests. The fraction of crystalline and amorphous domains may change upon temperature variations as a consequence of phase changes into the $\iota$-th polymer (*i.e.*, crystallization and melting). Particularly, for each polymer, it is possible to define and measure the characteristic temperatures of crystallization and melting, which well define the internal architecture of the material in terms of phase contents. These temperatures are denoted with $T_{c,\iota}$ and $T_{m,\iota}$, respectively, with $T_{m,\iota} > T_{c,\iota}$. Above $T_{m,\iota}$, the $\iota$-th polymer consists of only the amorphous domain and has a rubber-like behavior (*i.e.*, rubber elasticity); in proximity of $T_{c,\iota}$ the crystallites start to form, and once crystallization is completed, the $\iota$-th polymer is composed of crystallites connected by the amorphous cross-linked domain and has an elasto-plastic behavior. Upon subsequent heating, the formed crystallites start to melt up to the final restoration of the fully amorphous domain. Thus, in a shape-memory thermal cycle, the crystalline domain may be used as thermal switch and is responsible for both temporary shape fixation and permanent shape recovery.

The identification of all the melting and crystallization temperatures allows the definition of $T_m$ and $T_c$ for the whole network, as follows

$$T_m = \max \{T_{m,\iota}\}, \qquad T_c = \min \{T_{c,\iota}\}. \tag{1}$$

Accordingly, amorphous and crystalline phases of the network vary by means of cooling-heating cycles. For values of temperature above $T_m$, the entire multi-phase semi-crystalline network is in its fully amorphous state, with $n$ amorphous domains. The cooling process determines the sequential formation of the crystalline domains, which stiffen the material and reach their maximum contents below $T_c$. Therefore, for $T \leq T_c$, the multi-phase semi-crystalline network is made of $n$ crystalline domains, and $n$ amorphous domains. A schematic representation of the internal architecture evolution of the network, as the temperature varies, is depicted in Fig. 1.

It is clear the necessity to properly characterize the $n$ polymers, with associated amorphous and crystalline phase contents, to describe the overall internal architecture evolution of the material system as a function of temperature variations.

## 2.2. Phase variable definitions

Starting from the description of the material system in previous subsection, we denote with $\mathcal{V}$ the total volume of the entire network. It holds that

$$\mathcal{V} = \mathcal{V}_A + \mathcal{V}_C \,, \tag{2}$$

where the total volumes of the amorphous and crystalline phase contents, $\mathcal{V}_A$ and $\mathcal{V}_C$, are given as

$$\mathcal{V}_A = \sum_{\iota=1}^{n} \mathcal{V}_{\iota,A} \,, \qquad \mathcal{V}_C = \sum_{\iota=1}^{n} \mathcal{V}_{\iota,C} \,, \tag{3}$$

with $\mathcal{V}_{\iota,A}$ and $\mathcal{V}_{\iota,C}$ volumes of the amorphous and crystalline domain of any $\iota$-th polymer composing the network, respectively.

Similarly, it is possible to define the volume of the $\iota$-th polymer as

$$\mathcal{V}_\iota = \mathcal{V}_{\iota,A} + \mathcal{V}_{\iota,C} \,, \tag{4}$$

which satisfies Eq. (2) according to the relation $\mathcal{V} = \sum_\iota \mathcal{V}_\iota$.

Let $\chi$ be a scalar variable named total volume fraction. The amorphous and crystalline volume fractions of the entire network are denoted as $\chi_A$ and $\chi_C$ henceforth, and they are given as

$$\chi_A = \frac{\mathcal{V}_A}{\mathcal{V}} \,, \qquad \chi_C = \frac{\mathcal{V}_C}{\mathcal{V}} \,, \tag{5a}$$

from which descends that

$$\chi = \chi_A + \chi_C = 1 \,, \tag{5b}$$

according to Eq. (2). Similarly, each $\iota$-th polymer will have a total volume fraction $\chi_\iota = \mathcal{V}_\iota \mathcal{V}^{-1}$, as well as amorphous and crystalline volume fractions will be given as

$$\chi_{\iota,A} = \frac{\mathcal{V}_{\iota,A}}{\mathcal{V}} \,, \qquad \chi_{\iota,C} = \frac{\mathcal{V}_{\iota,C}}{\mathcal{V}} \,, \tag{6a}$$

whence

$$\chi_\iota = \chi_{\iota,A} + \chi_{\iota,C} \,, \tag{6b}$$

according to Eq. (4) and in respect of condition (5b). It therefore holds that

$$\chi = \sum_{\iota=1}^{n} \chi_\iota = 1 \,, \qquad \chi_A = \sum_{\iota=1}^{n} \chi_{\iota,A} \,, \qquad \chi_C = \sum_{\iota=1}^{n} \chi_{\iota,C}. \tag{7}$$

In compliance with the material system description provided in the previous subsection, it is reasonable to assume the following upper and lower bounds to any $\iota$-th volume fraction (6a)

$$\chi_{\iota,A}^{\min} < \chi_{\iota,A} \leq \chi_\iota \,, \qquad 0 \leq \chi_{\iota,C} \leq \chi_{\iota,C}^{\max} \,, \tag{8}$$

where $\chi_{\iota,A} = \chi_\iota$ and $\chi_{\iota,C} = 0$ hold for $T \geq T_{m,\iota}$, whereas $T \leq T_{c,\iota}$ implies that $\chi_{\iota,A} = \chi_{\iota,A}^{\min} = \chi_\iota - \chi_{\iota,C}^{\max} > 0$ and $\chi_{\iota,C} = \chi_{\iota,C}^{\max} < \chi_\iota$. Summation over $\iota$ yields

$$\chi_A^{\min} < \chi_A \leq 1 \,, \qquad 0 \leq \chi_C \leq \chi_C^{\max} \,, \tag{9}$$

which represent the upper and lower bounds of volume fractions (5a) that allow the overall characterization of the entire network, and where $\chi_A = 1$ and $\chi_C = 0$ for $T \geq T_m$, $\chi_A = \chi_A^{\min} = \sum_\iota \chi_{\iota,A}^{\min} > 0$ and $\chi_C = \chi_C^{\max} = \sum_\iota \chi_{\iota,C}^{\max} < 1$ for $T \leq T_c$.

In view of upper bounds (8), volume fractions $\chi_{\iota,C}$ can be conveniently selected to define the micro-structural phase evolution of any $\iota$-th polymer, which in turn drive the description of the phase contents (9) of the entire network. Hence, selecting all $\chi_{\iota,C}$ as independent phase variables, the amorphous volume fractions of the $\iota$-th polymer and of the entire network re-write in the form

$$\chi_{\iota,A} = \chi_\iota - \chi_{\iota,C} \,, \qquad \chi_A = 1 - \chi_C \,, \tag{10}$$

according to Eqs. (5b) and (6b).

## 2.3. Identification of the maximum crystallinity volume contents

Evaluation of the maximum crystalline volume contents, $\chi_{\iota,C}^{\max}$, in Eq. (8), yielding $\chi_C^{\max}$, can be provided via prior identification of the maximum crystalline weight contents, $\chi_{\iota,C}^{w,\max}$, which can be derived from experimental measurements performed by differential scanning calorimetry (Boatti et al., 2016; Scalet et al., 2018; Inverardi et al., 2022; Arricca et al., 2024).

To do that, let the symbols $\varrho$ and $\mathcal{M}$ denote the mass density and the mass of the entire network. For any given $\iota$-th polymer, it holds that

$$\varrho_{\iota,A} = \frac{\mathcal{M}_{\iota,A}}{\mathcal{V}_{\iota,A}} \,, \qquad \varrho_{\iota,C} = \frac{\mathcal{M}_{\iota,C}}{\mathcal{V}_{\iota,C}} \,, \tag{11}$$

with $\varrho_{\iota,A}$ and $\mathcal{M}_{\iota,A}$, $\varrho_{\iota,C}$ and $\mathcal{M}_{\iota,C}$ mass densities and masses of the amorphous and crystalline domains of any $\iota$-th polymer, respectively It follows that

$$\mathcal{M} = \sum_{\iota=1}^{n} \varrho_{\iota,A}\, \mathcal{V}_{\iota,A} + \varrho_{\iota,C}\, \mathcal{V}_{\iota,C}\,, \tag{12}$$

which provide a definition of the total mass of the network.

The generic $\iota$-th crystalline weight content is given as

$$\chi_{\iota,C}^{w} = \frac{\mathcal{M}_{\iota,C}}{\mathcal{M}}\,, \tag{13}$$

and its maximum value is obviously provided by the maximum value that the numerator can assume, *i.e.*, $\mathcal{M}_{\iota,C}^{\max} = \varrho_{\iota,C}\,\mathcal{V}_{\iota,C}^{\max}$ for $T \leq T_{c,\iota}$. Accordingly, and taking advantage of Eqs. (6), (11) and (12), the following relation applies to define the maximum crystalline weight content of any $\iota$-th polymer

$$\chi_{\iota,C}^{w,\max} = \frac{\varrho_{\iota,C}\,\chi_{\iota,C}^{\max}}{\varrho_{\iota,A}(\chi_{\iota} - \chi_{\iota,C}^{\max}) + \varrho_{\iota,C}\,\chi_{\iota,C}^{\max} + \sum_{j \neq \iota} \varrho_{j,\kappa}\,\chi_{j,\kappa}}\,, \tag{14}$$

for $\kappa = A, C$ and $j = 1, \ldots, n-1$.

Note that, although Eq. (14) is true for any given $\iota$-th polymer, the term $\sum_{j \neq \iota} \varrho_{j,\kappa}\,\chi_{j,\kappa}$ is not univocally defined as long as a fixed value of $T$ is not selected. Therefore, in order to state a meaningful phase content characterization of the network, consider $T \leq T_c$ such that the crystalline content of both the $\iota$-th polymer and all the remaining $n-1$ polymers have reached their maximum extent. Then, Eq. (14) re-writes in the form

$$\chi_{\iota,C}^{w,\max} = \frac{\varrho_{\iota,C}\,\chi_{\iota,C}^{\max}}{\varrho_{\iota,A}\,\chi_{\iota} + \chi_{\iota,C}^{\max}(\varrho_{\iota,C} - \varrho_{\iota,A}) + \sum_{j \neq \iota} \varrho_{j,A}\,\chi_{j} + \sum_{j \neq \iota} \chi_{j,C}^{\max}(\varrho_{j,C} - \varrho_{j,A})}\,, \tag{15}$$

whence the $\iota$-th maximum crystalline volume content writes as

$$\chi_{\iota,C}^{\max} = \chi_{\iota,C}^{w,\max}\, \frac{\varrho_{\iota,A}\,\chi_{\iota} + \sum_{j \neq \iota} \varrho_{j,C}\,\chi_{j} + \chi_{j,C}^{\max}(\varrho_{j,C} - \varrho_{j,A})}{\varrho_{\iota,C} - \chi_{\iota,C}^{w,\max}(\varrho_{\iota,C} - \varrho_{\iota,A})}\,. \tag{16}$$

Eq. (16) represents a system of $n$-equations in $n$-unknowns whose solution is

$$\chi_{\iota,C}^{\max} = \chi_{\iota,C}^{w,\max}\, \frac{\varrho_{\iota,A}\,\chi_{\iota} + \sum_{j \neq \iota} \varrho_{j,A}\,\chi_{j}}{\varrho_{\iota,C}\left[1 - \dfrac{\chi_{\iota,C}^{w,\max}(\varrho_{\iota,C} - \varrho_{\iota,A})}{\varrho_{\iota,C}} - \sum_{j \neq \iota} \dfrac{\chi_{j,C}^{w,\max}(\varrho_{j,C} - \varrho_{j,A})}{\varrho_{j,C}}\right]}\,. \tag{17}$$

Lastly, the mass density of the entire network, $\varrho = \mathcal{M}\,\mathcal{V}^{-1}$, can be expressed by means of the standard rule of mixture, *i.e.*,

$$\varrho = (1 - \chi_C)\,\varrho_A + \chi_C\,\varrho_C\,, \tag{18}$$

with $\varrho_A$ and $\varrho_C$ mass densities of the total amorphous and crystalline contents of the network, respectively. Here, advantage has been taken of relations $\mathcal{M} = \mathcal{M}_A + \mathcal{M}_C = \varrho_A\,\mathcal{V}_A + \varrho_C\,\mathcal{V}_C$, and of Eqs. (5a) and (10)$_2$.

## 3. Kinematics

This section presents the kinematic assumptions of the shape-memory material system under investigation.

The displacement of a particle from its initial position, $\mathbf{X} \in \mathcal{V}_0$, to the current position at time $t$, $\mathbf{x} \in \mathcal{V}_t$, with $\mathcal{V}_0$ and $\mathcal{V}_t$ initial (undeformed) and current (deformed) configuration, respectively, is defined by the deformation equation $\mathbf{x} = \mathbf{x}(\mathbf{X}, t)$. The deformation gradient referred to the undeformed configuration is denoted by $\mathbf{F}$ and is defined as $\mathbf{F} = \partial \mathbf{x}(\mathbf{X}, t)/\partial \mathbf{X}$ (Malvern, 1969), yielding a two-point tensor field $\mathbf{F} : \mathcal{V}_0 \to \mathcal{V}_t$.

### 3.1. Multiplicative decomposition of the total deformation gradient

The kinematic theory of the multi-phase semi-crystalline network under investigation is here assumed to be based on the following multiplicative decomposition of the deformation gradient

$$\mathbf{F} = \mathbf{F}^{m}\,\mathbf{F}^{th}\,, \tag{19}$$

where tensors $\mathbf{F}^{m}$ and $\mathbf{F}^{th}$ denote the mechanical and thermal deformation gradient, respectively. It holds that $\mathbf{F}^{th} : \mathcal{V}_0 \to \mathcal{V}_T$ and $\mathbf{F}^{m} : \mathcal{V}_T \to \mathcal{V}_t$, with $\mathcal{V}_T$ (intermediate) thermal configuration (see Fig. 2) at a non-uniform temperature $T$, which is obtained from the current configuration, $\mathcal{V}_t$, by isothermal de-stressing to zero stress (Lubarda, 2004).

**Remark.** It is perhaps worth to point out that configurations $\mathcal{V}_0$, $\mathcal{V}_T$, and $\mathcal{V}_t$ correspond to the total volume of the network (at different time, temperature, and stress), which we have simply denoted with $\mathcal{V}$ in Sections 2.2 and 2.3. The main aim of such sections was the introduction of the crystalline volume fractions as *phase* variables to describe the phase evolution of network, and their respective maximum contents. Volume fractions are obviously configurational invariants, so they do not require such specification, and all equations introduced in Sections 2.2 and 2.3 can be arbitrarily defined in any configurations.

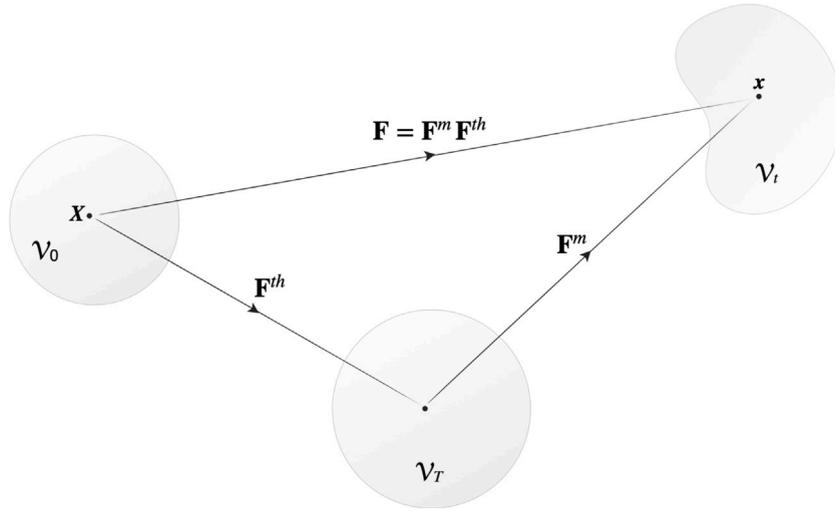

**Fig. 2.** Depiction of the initial (undeformed, or material) configuration $\mathcal{V}_0$, spatial (deformed, or current) configuration $\mathcal{V}_t$, and thermal (intermediate) configuration $\mathcal{V}_T$.

*3.2. Phase contributions to the mechanical deformation*

Standard arguments in SMP modeling commonly assume the equality between either stresses (Liu et al., 2006; Baghani et al., 2011, 2012; Scalet et al., 2018), or strains (Boatti et al., 2016; Yan et al., 2020; Arricca et al., 2024), between the two phases, therefore amorphous and crystalline in the present case. We here proceed according to the latter statement, keeping the thermal deformation gradient as total, therefore avoiding to provide kinematic specifications on the effects of thermal loads associated to the two phases. Accordingly, it is required that the mechanical contribution to the deformation gradient respects the following equality

$$\mathbf{F}^m = \mathbf{F}^m_A = \mathbf{F}^m_C. \tag{20}$$

with $\mathbf{F}^m_A$ and $\mathbf{F}^m_C$ mechanical deformation gradients in the amorphous and crystalline phases, respectively. In view of the thermal-sensitive behavior of the semi-crystalline SMP network, which varies from elastic-like for $T \geq T_m$ to elasto-plastic-like for $T \leq T_c$, tensors $\mathbf{F}^m_A$ and $\mathbf{F}^m_C$ are chosen to be given as

$$\mathbf{F}^m_A = \mathbf{F}^{el}_A, \qquad \mathbf{F}^m_C = \mathbf{F}^{el}_C \, \mathbf{F}^{in}. \tag{21}$$

Here, $\mathbf{F}^{el}_A$ represents the elastic deformation gradient in the amorphous phase, whereas $\mathbf{F}^{el}_C$ and $\mathbf{F}^{in}$ stand for the elastic and inelastic contributions to the deformation in the crystalline domain, and clarify the necessity to account for a mechanical deformation gradient in Eq. (19). It holds that $\mathbf{F}^{el}_A : \mathcal{V}_T \to \mathcal{V}_t$, $\mathbf{F}^{el}_C : \mathcal{V}_* \to \mathcal{V}_t$, and $\mathbf{F}^{in} : \mathcal{V}_T \to \mathcal{V}_*$, where $\mathcal{V}_*$ represents the inelastic configuration related to the crystalline content (see Fig. 3), which is obtained from $\mathcal{V}_t$ after removal of the elastic contribution to the stress.

Tensor $\mathbf{F}^{in}$ has to account for the deformations associated to the one-way and two-way SMEs taking place within the material. Accordingly, it can be subjected to a further multiplicative decomposition, as follows

$$\mathbf{F}^{in} = \mathbf{F}^{in}_C \, \mathbf{F}^{st,f}, \tag{22}$$

where $\mathbf{F}^{in}_C$ explicitly accounts for the elongations and contractions induced by the crystalline-phase evolution (known, respectively, as CIE and MIC, as discussed in Section 1), while $\mathbf{F}^{st,f}$ for deformations that can be temporary stored at low temperatures as a consequence of high-temperature loading (Boatti et al., 2016; Arricca et al., 2024). Combination of Eqs. (21)$_2$ and (22) would further lead to account for an intermediate configuration between $\mathcal{V}_T$ and $\mathcal{V}_*$, $\mathcal{V}_\S$ say, such that $\mathbf{F}^{st,f} : \mathcal{V}_T \to \mathcal{V}_\S$ and $\mathbf{F}^{in}_C : \mathcal{V}_\S \to \mathcal{V}_*$.

Note that, the $n$ polymers composing the network will contribute to the inelastic deformation as

$$\mathbf{F}^{in}_C = \prod_{i=1}^{n} \mathbf{F}^{in}_{i,C}, \qquad \mathbf{F}^{st,f} = \prod_{i=1}^{n} \mathbf{F}^{st,f}_i, \tag{23}$$

whereas the elastic contributions, although we may assume to be able to state the relations $\mathbf{F}^{el}_A = \prod_i \mathbf{F}^{el}_{i,A}$ and $\mathbf{F}^{el}_C = \prod_i^n \mathbf{F}^{el}_{i,C}$, require to be kept total, as we will made clear in Section 4.1.

**4. Constitutive equations**

Denote with $\psi$ the specific Helmholtz free energy (per unit mass) of the entire network, and with $\psi_A$ and $\psi_C$ the specific energies of the amorphous and crystalline phase contents. Within the model of the multiplicative decomposition (19), energy $\psi$ is commonly split into two parts, encompassing a strain energy function, $\psi^{el}$ say, and a thermal contribution, $\psi^{th}$, yielding $\psi = \psi^{el} + \psi^{th}$. With the aim of characterizing the constitutive stress–strain response of the system, we here focus on the only contribution $\psi^{el}$, therefore avoiding the treatment of the thermal energy, which is commonly written as a function of temperature and heat capacity and does not generate additional contributions to the stress in the material.

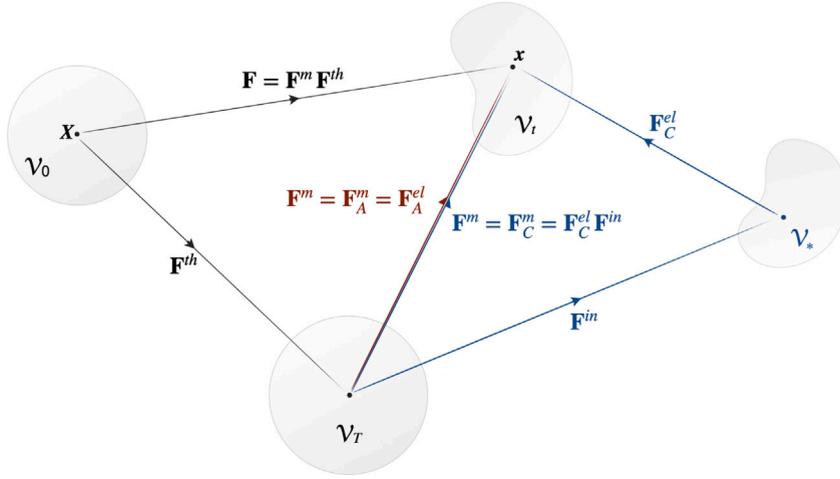

**Fig. 3.** Initial, thermal, and spatial configurations, with specifications on the mechanical contribution to the deformation in the amorphous and crystalline domains. Accounting for thermal loads, $\mathbf{F}_A^m = \mathbf{F}_A^{el}$ leads to the configurational picture of standard thermo-elasticity for the amorphous content. Rather, $\mathbf{F}_C^m = \mathbf{F}_C^{el} \mathbf{F}^{in}$ generates the intermediate inelastic configuration, $\mathcal{V}_*$, yielding a thermo-elasto-plastic scenario for the crystalline phase.

We can write the elastic contribution to the Helmholtz free energy according to the rule of mixture, *viz.*,

$$\psi^{el} = \chi_A^w \psi_A^{el} + \chi_C^w \psi_C^{el} = \frac{1}{\varrho} \left[ (1-\chi_C) \varrho_A \psi_A^{el} + \chi_C \varrho_C \psi_C^{el} \right], \qquad (24)$$

with $\psi_A^{el}$ and $\psi_C^{el}$ specific strain energies of the amorphous and crystalline domains. Here, advantage has been taken of definitions of amorphous and crystalline weight contents and of mass densities, $\chi_\kappa^w = \mathcal{M}_\kappa \mathcal{M}^{-1}$ and $\varrho_\kappa = \mathcal{M}_\kappa \mathcal{V}_\kappa^{-1}$, for $\kappa = A, C$, and of volume fractions (5a). Note that Eq. (24) is configurational invariant, and for the sake of generality we have avoided specifications on the configuration where mass densities lie.

For isotropic materials, and to satisfy the principle of objectivity of material properties (Noll, 1958), $\psi_A^{el}$ and $\psi_C^{el}$ will be functions of the invariants of the Green–Lagrange strain or, equivalently, the right or left stretch tensors (Malvern, 1969; McMeeking and Landis, 2005). We introduce the finite logarithmic strain-based extension of the standard linear elastic material (Hencky, 1933; de Souza Neto et al., 2008), to provide the Eulerian description of the strain energy density (per unit current volume).

Consider the polar decomposition of the elastic deformation gradients of the amorphous and crystalline volume contents,

$$\mathbf{F}_\kappa^{el} = \mathbf{R}_\kappa^{el} \mathbf{U}_\kappa^{el} = \mathbf{V}_\kappa^{el} \mathbf{R}_\kappa^{el}, \qquad (25)$$

for $\kappa = A, C$. Tensor $\mathbf{R}_A^{el}$ ($\mathbf{R}_C^{el}$) represents a pure rotation between $\mathcal{V}_T$ ($\mathcal{V}_*$) and $\mathcal{V}_t$, whereas $\mathbf{U}_A^{el}$ ($\mathbf{U}_C^{el}$) and $\mathbf{V}_A^{el}$ ($\mathbf{V}_C^{el}$) denote the right and left elastic stretch tensors in the amorphous (crystalline) domain. Stretch tensors are linked to the right and left elastic Cauchy–Green strains according to the following relations

$$\mathbf{U}_\kappa^{el} = \sqrt{\mathbf{F}_\kappa^{el\,T} \mathbf{F}_\kappa^{el}}, \qquad \mathbf{V}_\kappa^{el} = \sqrt{\mathbf{F}_\kappa^{el} \mathbf{F}_\kappa^{el\,T}}. \qquad (26)$$

The logarithmic elastic strains, $\varepsilon_A^{el} : \mathcal{V}_t \to \mathcal{V}_t$ and $\varepsilon_C^{el} : \mathcal{V}_t \to \mathcal{V}_t$, can be defined as a function of the left elastic stretch tensors (26)$_2$ as

$$\varepsilon_A^{el} = \ln \mathbf{V}_A^{el} = \ln \sqrt{\mathbf{F}_A^{el} \mathbf{F}_A^{el\,T}}, \qquad \varepsilon_C^{el} = \ln \mathbf{V}_C^{el} = \ln \sqrt{\mathbf{F}_C^{el} \mathbf{F}_C^{el\,T}}, \qquad (27)$$

where $\ln(*)$ denotes the tensor logarithm of $(*)$. Tensors $\varepsilon_A^{el}$ and $\varepsilon_C^{el}$ are selected as strain internal variables of energy (24), which can be written in terms of spatial energy density (per unit current volume) as

$$\varrho_t \psi^{el}(T, \varepsilon_A^{el}, \varepsilon_C^{el}) = (1-\chi_C) \varrho_{A_t} \psi_A^{el}(T, \varepsilon_A^{el}) + \chi_C \varrho_{C_t} \psi_C^{el}(T, \varepsilon_C^{el}). \qquad (28)$$

Taking advantage of the Hencky strain–energy relation, we provide the following definitions of the elastic energy densities of the amorphous and crystalline content

$$\varrho_{A_t} \psi_A^{el}(T, \varepsilon_A^{el}) = \frac{1}{2} \varepsilon_A^{el} : \mathbb{C}_{A_t}(T) \, \varepsilon_A^{el}, \qquad \varrho_{C_t} \psi_C^{el}(T, \varepsilon_C^{el}) = \frac{1}{2} \varepsilon_C^{el} : \mathbb{C}_{C_t} \, \varepsilon_C^{el}. \qquad (29)$$

Here, $\mathbb{C}_{A_t}$ and $\mathbb{C}_{C_t}$ have the same format of the infinitesimal fourth-order elasticity tensor, and material parameters related to the amorphous and crystalline domain content are taken as temperature-dependent and temperature-independent, respectively. Combination of Eqs. (28) and (29) yields

$$\varrho_t \psi^{el}(T, \varepsilon_A^{el}, \varepsilon_C^{el}) = \frac{1}{2} \left[ (1 - \chi_C) \varepsilon_A^{el} : \mathbb{C}_{A_t}(T) \varepsilon_A^{el} + \chi_C \varepsilon_C^{el} : \mathbb{C}_{C_t} \varepsilon_C^{el} \right]. \tag{30}$$

Note that, the logarithm of the left stretch does not, in general, have a conjugate stress. However, in the case of isotropic elastic materials, its conjugate stress is the Cauchy stress multiplied by the third principal invariant of the left stretch (Hoger, 1987), which corresponds to the Kirchhoff stress. Let $\boldsymbol{\sigma}_A : \mathcal{V}_t \to \mathcal{V}_t$ and $\boldsymbol{\sigma}_C : \mathcal{V}_t \to \mathcal{V}_t$, $\mathcal{I}_3(\mathbf{V}_A^{el})$ and $\mathcal{I}_3(\mathbf{V}_C^{el})$, be the Cauchy (true) stress tensors, and the third principal invariants of elastic left stretch tensors, in the amorphous and crystalline domains, respectively. Therefore, the stresses conjugate to logarithmic strains (27) write as

$$\boldsymbol{\sigma}_A^{[V]} = \mathcal{I}_3(\mathbf{V}_A^{el}) \boldsymbol{\sigma}_A, \qquad \boldsymbol{\sigma}_C^{[V]} = \mathcal{I}_3(\mathbf{V}_C^{el}) \boldsymbol{\sigma}_C. \tag{31}$$

Note that, although the Kirchhoff stress preserves the same mapping properties of the Cauchy stress, hence, $\boldsymbol{\sigma}_A^{[V]} : \mathcal{V}_t \to \mathcal{V}_t$ and $\boldsymbol{\sigma}_C^{[V]} : \mathcal{V}_t \to \mathcal{V}_t$, in view of the mapping properties of $\mathbf{F}_A^{el}$ and $\mathbf{F}_C^{el}$, Eqs. (31)$_1$ and (31)$_2$ have the measure of force per unit thermal and inelastic volume, respectively. However, accounting for the standard assumption of plastic incompressibility, also $\boldsymbol{\sigma}_C^{[V]}$ results measured per unit thermal volume, as the mass density of the crystalline domain is preserved from $\mathcal{V}_*$ to $\mathcal{V}_T$. Accordingly, the following constitutive stress–strain relations can be stated via derivation of Eqs. (29)

$$\boldsymbol{\sigma}_A^{[V]} = \varrho_{A_T} \frac{\partial \psi_A^{el}}{\partial \varepsilon_A^{el}} = \mathbb{C}_{A_T}(T) \varepsilon_A^{el}, \qquad \boldsymbol{\sigma}_C^{[V]} = \varrho_{C_T} \frac{\partial \psi_C^{el}}{\partial \varepsilon_C^{el}} = \mathbb{C}_{C_T} \varepsilon_C^{el}, \tag{32}$$

with $\varrho_{A_T} = \mathcal{I}_3(\mathbf{V}_A^{el}) \varrho_{A_t}$ and $\varrho_{C_T} = \mathcal{I}_3(\mathbf{V}_C^{el}) \varrho_{C_t}$, and consequent re-definition of material moduli of the amorphous and crystalline domains in force per unit thermal volume. The following definition of total (Kirchhoff) stress in the material can be eventually stated

$$\boldsymbol{\sigma}^{[V]} = (1 - \chi_C) \boldsymbol{\sigma}_A^{[V]} + \chi_C \boldsymbol{\sigma}_C^{[V]}. \tag{33}$$

**Remark.** Note that the fourth-order elasticity tensors introduced in Eqs. (29) are not necessarily isotropic. For the sake of generality, it is perhaps reasonable to assume the isotropy of tensor $\mathbb{C}_{A_t}$, whereas crystalline domains may present some directional dependences of their material properties, yielding $\mathbb{C}_{C_t}$ a non-isotropic tensor. In the following, the proposed model is validated for a case-study in Section 5. In view of the experimental data available, for the sake of simplicity both $\mathbb{C}_{A_t}$ and $\mathbb{C}_{C_t}$ will be taken as isotropic tensors.

### 4.1. Insights on the free energy identification

It is noteworthy to remark the non-trivial, and non-unique, identification of the strain energy function of the network. According to the selected Hencky model, the total energy of the system should be $\varrho_t \psi^{el} = \frac{1}{2} \varepsilon^{el} : \mathbb{C}_t \varepsilon^{el}$. However, in view of kinematic specifications (20) and (21), a straightforward definition of the total strain energy is not allowed, as it would require the identification of a total elastic strain tensor, $\varepsilon^{el} = \ln \mathbf{V}^{el}$, with $\mathbf{V}^{el} = \sqrt{\mathbf{F}^{el} \mathbf{F}^{el\mathrm{T}}}$, hence of $\mathbf{F}^{el}$, which cannot be univocally defined. Therefore, the strain energy of the network is stated as a result of the primary definition of the energies of the amorphous and crystalline domains, as the selected strain internal variables suggest, and as performed prior to the introduction of Eq. (30). In turn, the same applies to the total stress in the material (33), which we had to introduce after identification of stresses (32).

Note however, that kinematic specifications (21) allows us to state the different behavior of the two phases, which actually explore different deformations and different stresses, whereas assumption (20) has the aim to provide a useful tool that facilitates the analytical characterization of the stress–strain response of the material.

A further consideration can be made in terms of strain energy of any $\iota$-th polymer of the network for $n \geq 2$, and establishment of consequent stress–strain constitutive relations. That is, the identification of individual $\mathbf{F}_\iota^{el}$, which would allow us to define the specific strain energy $\psi_\iota^{el}$, is prevented by the impossibility to clearly measure the individual elastic deformation that each polymer experiences when embedded in a multi-phase network. However, if pursuing an alternative procedure we allowed the identification of $\mathbf{F}_\iota^{el}$, yielding $\mathbf{V}_\iota^{el}$ and the definition of $\psi_\iota^{el}$, the energy of the system would be simply given as $\sum_\iota \psi_\iota^{el}$. Hence, it would result from a trivial summation of all $n$ energies without the possibility to perform any averaging procedures, therefore to account for interactions among polymers. Alternatively, if the definition of the individual strain energies of the amorphous and crystalline domains, $\psi_{\iota,A}^{el}$ and $\psi_{\iota,C}^{el}$, was allowed via identification of tensors $\mathbf{F}_{\iota,A}^{el}$ and $\mathbf{F}_{\iota,C}^{el}$, the energy of the amorphous and crystalline domains would be $\sum_\iota \psi_{\iota,A}^{el}$ and $\sum_\iota \psi_{\iota,C}^{el}$. In this case the overall behavior of the network would result even less plausible, as provided by individual and non-interacting contributions arising from each phase of each polymer. Furthermore, unrealistic statements of any material moduli of each $\iota$-th polymer, or of any material moduli of each $\iota$-th amorphous and crystalline domain, would be necessary.

For the aforementioned reasons, we have required that the elastic deformation of the amorphous and crystalline domains must be kept total, as specified in Section 3.2, allowing the selection of the strain internal variables (27) of energies (29) and definitions of stresses (32). Such procedure probably represents the most reasonable approach to characterize the strain energy of the system, and the stress–strain relations. In fact, partial mutual interactions among the components within the network are implicitly accounted for in view of the weighted procedure performed by volume fractions (7)$_2$ and (7)$_3$ on the total energies of the amorphous and crystalline domains.

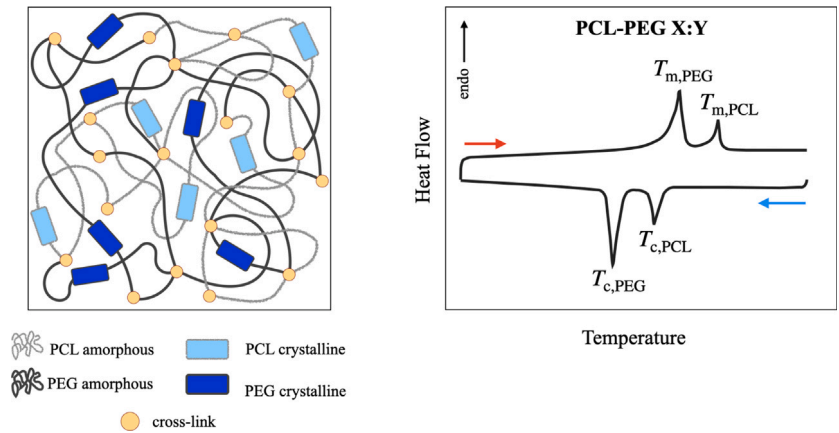

**Fig. 4.** Schematic of the internal architecture of the PCL-PEG network, below the crystallization temperature, used for model validation (left) and of DSC cooling and second heating scans of a generic PCL-PEG weight ratio (right).

## 5. Case-study: application to copolymer networks

This section applies the general formulation presented in previous sections to the case of a two-phase semi-crystalline network to validate the model.

Specifically, some poly(ethylene glycol) (PEG)/poly($\varepsilon$-caprolactone) (PCL) semi-crystalline networks synthesized and experimentally characterized under uniaxial conditions by Inverardi et al. (2022) are considered. They were prepared by photo-cross-linking of methacrylated macromonomers with different molecular weights and weight ratios. Particularly, the systems obtained using PCL diol ($M_n \sim 10$kDa, PCL10) and PEG with average molecular weight of $M_n \sim 3$ kDa (PEG3) at different weight ratios of 1:2, 1:1, and 2:1 were used for validation.

It is noted that the choice of focusing on PCL-PEG-based systems does not restrain the generality of the proposed model and its suitable application to any multi-phase semi-crystalline cross-linked networks exhibiting the one-way SME as well as the two-way SME under stress or stress-free conditions.

### 5.1. Phase variables

Following the adopted nomenclature and notation, the network is a two-phase semi-crystalline network composed of $n = 2$ polymers (*i.e.,* PCL10 and PEG3), and thus of two amorphous domains, and two crystalline domains for values of temperature below the network crystallization temperature, as schematically depicted in Fig. 4 (left). For the sake of notation simplicity, the two polymers of the network are denoted with the labels PCL and PEG hereinafter. Therefore, symbol $\iota = $ PCL, PEG.

Accordingly, the crystalline volume fractions of the PCL and PEG polymers can be represented by introducing two independent phase variables denoted with $\chi_{\text{PCL},C}$ and $\chi_{\text{PEG},C}$, respectively. The adopted values for $\chi_{\text{PCL}}$ and $\chi_{\text{PEG}}$ are properly selected as a function of the three different weight ratios that define the three different proportions in which the two polymers constitute the network and are reported in Table 1.

The crystalline and amorphous domains have distinct crystallization and melting temperatures, as shown in a representative differential scanning calorimetry (DSC) curve for a generic PCL-PEG weight ratio in Fig. 4 (right). The measured temperatures for the three PCL-PEG-based systems under investigation are taken from Inverardi et al. (2022) and are listed in Table 1. Note however, that some of the values of the crystalline weight contents are missing in Inverardi et al. (2022), as well as some of the crystallization and melting temperatures adopted here. In the current work, further *ad-hoc* DSC tests have been performed in order to provide all the necessary information.

According to Eq. (1) and experimental measurements, the following melting and crystallization temperatures of the entire network are thus identified

$$T_m = T_{m,\text{PCL}}, \qquad T_c = T_{c,\text{PEG}}. \tag{34}$$

Therefore, the upper and lower bounds of the total amorphous and crystalline volume fractions (9) respect the conditions $\chi_A = 1$ and $\chi_C = 0$ for $T \geq T_{m,\text{PCL}}$, $\chi_A = \chi_A^{\min}$ and $\chi_C = \chi_C^{\max}$ for $T \leq T_{c,\text{PEG}}$, with

$$\chi_A = \chi_{\text{PCL},A} + \chi_{\text{PEG},A}, \qquad \chi_C = \chi_{\text{PCL},C} + \chi_{\text{PEG},C}. \tag{35}$$

**Table 1**
Crystallization and melting temperatures as well as crystallinity weight contents measured through DSC tests by Inverardi et al. (2022) and through *ad-hoc* DSC tests performed in this work for the PEG-PCL-based systems under investigation.

| Parameter | PCL-PEG 2:1 | PCL-PEG 1:1 | PCL-PEG 1:2 | Units |
| --- | --- | --- | --- | --- |
| $\chi_{PEG}$ | 1/3 | 1/2 | 2/3 | – |
| $\chi_{PCL}$ | 2/3 | 1/2 | 1/3 | – |
| $T_{c,PCL}$ | 18 | 18 | 21 | °C |
| $T_{m,PCL}$ | 44 | 43 | 44 | °C |
| $T_{c,PEG}$ | –1 | –7 | –5 | °C |
| $T_{m,PEG}$ | 24 | 18 | 19 | °C |
| $\chi_{PCL,C}^{w,max}$ | 0.18 | 0.14 | 0.17 | – |
| $\chi_{PEG,C}^{w,max}$ | 0.12 | 0.30 | 0.23 | – |

The values of the maximum crystallinity weight contents, $\chi_{PCL,C}^{w,max}$ and $\chi_{PEG,C}^{w,max}$, whose evaluation is carried out via DCS tests, are reported in Table 1. For $n=2$, $\iota, J = \text{PCL, PEG}$ with $\iota \neq J$, and values of temperature below $T_{c,PEG}$, Eq. (17) allows the evaluation of the maximum crystallinity volume contents as

$$\chi_{\iota,C}^{max} = \chi_{\iota,C}^{w,max} \frac{\varrho_{\iota,A}\chi_{\iota} + \varrho_{J,A}\chi_{J}}{\varrho_{\iota,C}\left[1 - \frac{\chi_{\iota,C}^{w,max}(\varrho_{\iota,C}-\varrho_{\iota,A})}{\varrho_{\iota,C}} - \frac{\chi_{J,C}^{w,max}(\varrho_{J,C}-\varrho_{J,A})}{\varrho_{J,C}}\right]}. \quad (36)$$

The relation to evaluate the crystallinity weight contents of PCL and PEG, which has an analogous form of Eq. (15), is given as

$$\chi_{\iota,C}^{w} = \frac{\varrho_{\iota,C}\chi_{\iota,C}}{\varrho_{\iota,A}\chi_{\iota} + \chi_{\iota,C}(\varrho_{\iota,C}-\varrho_{\iota,A}) + \varrho_{J,A}\chi_{J} + \chi_{J,C}(\varrho_{J,C}-\varrho_{J,A})}. \quad (37)$$

The PCL amorphous and crystalline mass densities are taken as $\varrho_{PCL,A} = 1.081\,\text{g cm}^{-3}$ and $\varrho_{PCL,C} = 1.195\,\text{g cm}^{-3}$ (Ketelaars et al., 1997), whereas data used for PEG are $\varrho_{PEG,A} = 1.064\,\text{g cm}^{-3}$ and $\varrho_{PEG,C} = 1.2\,\text{g cm}^{-3}$ (Martuscelli et al., 1986).

### 5.2. Logarithmic stretch-based kinematics

Consider a tensile test where the specimen does not undergo any shear deformation, so that the total deformation gradient have components $F_{11} = \lambda$, $F_{22} = F_{33} = \lambda_\perp$, $F_{ij} = 0$ for $i \neq j$. Therefore, we restrict the analysis in the sequel to non-rotative systems, where all deformation gradients arising from multiplicative decompositions performed in Section 3 are diagonal two-point tensors. Term $F_{11}$ represents the deformation along which the force is applied, where the stress has the only non-zero component. In the current numerical example we evaluate only the (scalar) stress and the (scalar) strain along the direction of force application, for which experimental data are available (Inverardi et al., 2022). For the sake of simplicity, subscripts$_{11}$ are omitted to avoid redundant notations.

Kinematic specifications introduced in Section 3 can be re-defined in terms of stretches after identification of lengths $\ell_0$, $\ell_T$, and $\ell_t$, oriented along the direction of force application, within configurations $\mathcal{V}_0$, $\mathcal{V}_T$ and $\mathcal{V}_t$, respectively. The axial stretch is given by $\lambda = \ell_t \ell_0^{-1}$. Upon mechanical unloading from $\ell_t$, the specimen will have length $\ell_T$, therefore defining the thermal stretch associated to the initial configuration as $\lambda^{th} = \ell_T \ell_0^{-1}$. Since the unloading from $\ell_t$ to $\ell_T$ is a mechanical process, we are justified to introduce the mechanical stretch as $\lambda^m = \ell_t \ell_T^{-1}$ (de Souza Neto et al., 2008). Accordingly, any deformed state of the network is characterized by the following multiplicative split of the axial stretch

$$\lambda = \lambda^m \lambda^{th}, \quad (38)$$

which corresponds to the multiplicative decomposition of the deformation gradient (19).

Within this setting, allowance for the strain-stretch relation is made for all deformations. Therefore, from Eq. (38) we get $\varepsilon = \ln \lambda$, and the following additive decomposition of strains holds

$$\varepsilon = \varepsilon^m + \varepsilon^{th}, \quad (39)$$

with $\varepsilon^m = \ln \lambda^m$ and $\varepsilon^{th} = \ln \lambda^{th}$ mechanical and thermal axial strain, respectively, and where the former yields

$$\varepsilon^m = \varepsilon_A^m = \varepsilon_C^m, \qquad \varepsilon_A^m = \varepsilon_A^{el}, \qquad \varepsilon_C^m = \varepsilon_C^{el} + \varepsilon^{in}, \quad (40)$$

after re-arrangement of kinematic specifications (20) and (21) in terms of stretches, and being $\varepsilon_A^{el} = \ln \lambda_A^{el}$, $\varepsilon_C^{el} = \ln \lambda_C^{el}$, and $\varepsilon^{in} = \ln \lambda^{in}$. According to Fig. 3, elastic and inelastic stretches are defined as $\lambda_A^{el} = \ell_t \ell_T^{-1}$, $\lambda_C^{el} = \ell_t \ell_*^{-1}$, and $\lambda^{in} = \ell_* \ell_T^{-1}$, with $\ell_*$ length oriented along the direction of force application within $\mathcal{V}_*$. The logarithmic stretch-based kinematics introduced in the current section has the same format of the *Voigt* model, as depicted in Fig. 5.

Accounting for Eq. (22) further allows us to re-write the inelastic strain as

$$\varepsilon^{in} = \varepsilon_C^{in} + \varepsilon^{st,f}, \quad (41)$$

where it holds that $\varepsilon_C^{in} = \ln \lambda_C^{in}$ and $\varepsilon^{st,f} = \ln \lambda^{st,f}$, $\lambda_C^{in} = \ell_* \ell_\S^{-1}$ and $\lambda_\S^{st,f} = \ell_t \ell_T^{-1}$ with the obvious meaning of stretches and lengths.

According to specifications provided in Section 4, and the selected strain internal variables of the total energy (28) via energy functions (29), elastic strains associated with the amorphous and crystalline phase are kept total. Rather, following Eq. (23), the inelastic response will be characterized by stating proper evolution laws to describe $\varepsilon_{\text{PCL}}^{in}$ and $\varepsilon_{\text{PEG}}^{in}$ separately. Therefore, use of Eq. (40)$_2$ has to be made to define the elastic deformation in the amorphous volume content, whereas the crystalline mechanical strain in Eq. (40)$_3$ needs to be taken as

$$\varepsilon_C^m = \varepsilon_C^{el} + \varepsilon_{\text{PCL}}^{in} + \varepsilon_{\text{PEG}}^{in}, \tag{42}$$

where the last two right-hand side terms are given as

$$\varepsilon_{\text{PCL}}^{in} = \varepsilon_{\text{PCL},C}^{in} + \varepsilon_{\text{PCL}}^{st,f}, \qquad \varepsilon_{\text{PEG}}^{in} = \varepsilon_{\text{PEG},C}^{in} + \varepsilon_{\text{PEG}}^{st,f}, \tag{43}$$

according to Eqs. (23).

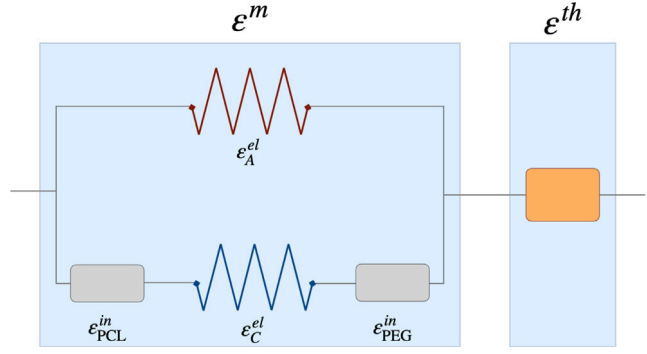

**Fig. 5.** Schematic of the additive decomposition of the total strain for the PCL-PEG system under investigation.

As specified in Section 3.2 for deformation gradients, strain $\varepsilon_{\text{PCL},C}^{in}$ ($\varepsilon_{\text{PEG},C}^{in}$) accounts for the cooling-induced elongations and the melting-induced contractions associated with the PCL (PEG) crystallization evolution, whereas $\varepsilon_{\text{PCL}}^{st,f}$ ($\varepsilon_{\text{PEG}}^{st,f}$) describes the deformation induced by high-temperature loading that is temporarily stored at low temperature during the cooling cycle, and released in full after heating above $T \geq T_{\text{m,PCL}}$ ($T \geq T_{\text{m,PEG}}$) (Boatti et al., 2016).

**Remark.** It is noteworthy to remark, that switching from the multiplicative decomposition of deformation gradients to the additive decomposition of logarithmic strain-based deformation tensors is not always allowed and some clarifications must be made. In fact, starting from Eq. (38), its tensorial form definition writes as $\boldsymbol{\varepsilon} = \boldsymbol{\varepsilon}^m + \boldsymbol{\varepsilon}^{th}$, which arises by selecting $\boldsymbol{\varepsilon} = \ln \mathbf{V}$, yielding $\boldsymbol{\varepsilon}^m = \ln \mathbf{V}^m$ and $\boldsymbol{\varepsilon}^{th} = \ln \mathbf{V}^{th}$, with $\mathbf{V}$, $\mathbf{V}^m$ and $\mathbf{V}^{th}$, total, mechanical and thermal left stretch tensor, respectively. Following the same path of reasoning, analogous tensorial definitions can be stated for Eqs. (40)–(43). However, such procedure needs to be restricted to the case of symmetric deformations, where therefore left and right stretch tensors are coincident, as each deformation gradient is the same as its transpose (see Eqs. (25) and (26), for instance). In fact, without loss of generality, consider two generic tensors, $\mathbf{Y}$ and $\mathbf{Z}$. Then, $\ln \mathbf{YZ} = \ln \mathbf{Y} + \ln \mathbf{Z} = \ln \mathbf{ZY}$ implies that $\mathbf{YZ} = \mathbf{ZY}$, therefore requiring that $\mathbf{Y}, \mathbf{X} \in Sym$.

Therefore, allowance for such procedure is made for our case of study.

### 5.3. Constitutive stress–strain response

In view of the selected Hencky strain–energy relations (29) and the consequent stress–strain responses (32), the stresses in the amorphous and crystalline domains are given as

$$\sigma_A^{[V]} = E_A(T)\,\varepsilon_A^{el} = E_A(T)\,\varepsilon^m, \qquad \sigma_C^{[V]} = E_C\,\varepsilon_C^{el} = E_C\,(\varepsilon^m - \varepsilon^{in}), \tag{44}$$

where advantage has been taken of Eqs. (40) to replace the elastic axial strains. $E_A(T)$ and $E_C$ represent the temperature-dependent and -independent elastic moduli of the amorphous and crystalline phases of the entire network, respectively. The former is taken as linearly dependent on the temperature according to the relation $E_A(T) = E_A' T$ (Scalet et al., 2018; Arricca et al., 2024), being $E_A'$ a positive elastic modulus.

According to Eqs. (44), the total stress in the material (33) writes as

$$\sigma^{[V]} = (1 - \chi_C)\,E_A' T\,\varepsilon^m + \chi_C\,E_C\,(\varepsilon^m - \varepsilon^{in}), \tag{45}$$

with $\chi_C = \chi_{\text{PCL},C} + \chi_{\text{PEG},C}$, according to Eq. (7)$_3$, and

$$\varepsilon^{in} = \varepsilon_{\text{PCL}}^{in} + \varepsilon_{\text{PEG}}^{in} = \varepsilon_{\text{PCL},C}^{in} + \varepsilon_{\text{PCL}}^{st,f} + \varepsilon_{\text{PEG},C}^{in} + \varepsilon_{\text{PEG}}^{st,f}, \tag{46}$$

in view of Eqs. (43).

Introduction of the total elastic modulus of the PCL-PEG network, $E$, as (Scalet et al., 2018; Arricca et al., 2024)

$$E = (1 - \chi_C)\,E_A' T + \chi_C\,E_C, \tag{47}$$

allows the following re-definition of stress (45)

$$\sigma^{[V]} = E\,\varepsilon^m - \chi_C\,E_C\,\left(\varepsilon^{in}_{\text{PCL}} + \varepsilon^{in}_{\text{PEG}}\right) = E\,(\varepsilon - \varepsilon^{th}) - \chi_C\,\sigma^{[V]}_{act}, \tag{48}$$

after replacement of the mechanical strain via Eq. (39). Term $\sigma^{[V]}_{act}$ denotes the stress associated with the actuation process under stress-free conditions, which is given as

$$\sigma^{[V]}_{act} = E_C\,\left(\varepsilon^{in}_{\text{PCL}} + \varepsilon^{in}_{\text{PEG}}\right) = E_C\,\left(\varepsilon^{in}_{\text{PCL},C} + \varepsilon^{st,f}_{\text{PCL}} + \varepsilon^{in}_{\text{PEG},C} + \varepsilon^{st,f}_{\text{PEG}}\right), \tag{49}$$

in view Eqs. (43). Eq. (48) further allows us to describe the total strain in the network as

$$\varepsilon = \varepsilon^{th} + \frac{1}{E}(\sigma^{[V]} + \chi_C\,\sigma^{[V]}_{act}), \tag{50}$$

where strain $\varepsilon^{th}$ is identified by means of the following definition of thermal stretch

$$\lambda^{th} = 1 + \left[(1-\chi_C)\,\alpha_A + \chi_C\,\alpha_C\right]\,\Delta T - \chi_C\,\Delta\varepsilon^{th}_C. \tag{51}$$

Here, $\alpha_A$ and $\alpha_C$ denote the thermal expansion coefficients of the amorphous and crystalline domains of the entire network, $\Delta T = T - T_0$, with $T_0$ initial temperature, and $\Delta\varepsilon^{th}_C$ represents the strain increment parameter related to the elongations and contraction occurring during crystallization and melting, respectively (Scalet et al., 2018; Arricca et al., 2024). Eq. (51) has been stated starting from the approximated expression $\lambda^{th} \simeq 1 + \alpha\,\Delta T$ (Vujošević and Lubarda, 2002), and has been modified by modeling the thermal expansion coefficient, $\alpha$, via the rule of mixture, and augmented by means of the last right-hand side term to account for CIE and MIC processes.

Given the constitutive relation of the total stress in the material (48), we eventually derive the total Piola (nominal) stress, P say. Consider $A_0$ and $\ell_0$ as the cross-sectional area and the length of the undeformed specimen, which are deformed to $A_t$ and $\ell_t$ after application of the uniaxial force $f$. Assuming that the SMP network under investigation is a nearly incompressible system, Kirchhoff stresses coincide with Cauchy stresses. Therefore, writing the Cauchy and Piola stresses as $\sigma = f\,A_t^{-1}$ and $P = f\,A_0^{-1}$, respectively, and setting $A_0\,\ell_0 \simeq A_t\,\ell_t$, it holds that

$$P = \sigma\,\lambda^{-1} = \sigma\,\lambda^{-m}\,\lambda^{-th}, \tag{52}$$

in accordance with multiplicative split (38).

### 5.4. Evolution laws

The temperature-dependent evolution of the crystalline phase within a cooling and subsequent heating cycle is evaluated in terms of volume content and given as

$$\dot{\chi}_C = \dot{\chi}_{\text{PCL},C} + \dot{\chi}_{\text{PEG},C}. \tag{53}$$

Set $\kappa = \text{PCL}, \text{PEG}$. We state that (Arricca et al., 2024)

$$\dot{\chi}_{\kappa,C} = -\begin{cases} \dfrac{\zeta^{\text{cool}}_\kappa \exp\left[\zeta^{\text{cool}}_\kappa (T - T^{\text{eff}}_{c,\kappa})\right]}{\left(1 + \exp\left[\zeta^{\text{cool}}_\kappa (T - T^{\text{eff}}_{c,\kappa})\right]\right)^2}\,\chi^0_{\kappa,C}\,\dot{T}, & \dot{T} \leq 0, \\[6pt] \dfrac{\zeta^{\text{heat}}_\kappa \exp\left[\zeta^{\text{heat}}_\kappa (T - T^{\text{eff}}_{m,\kappa})\right]}{\left(1 + \exp\left[\zeta^{\text{heat}}_\kappa (T - T^{\text{eff}}_{m,\kappa})\right]\right)^2}\,\chi^0_{\kappa,C}\,\dot{T}, & \dot{T} > 0, \end{cases} \tag{54}$$

which allows the description of the cyclic transformation of $\chi_{\text{PCL},C}$ and $\chi_{\text{PEG},C}$ within multiple cooling-heating steps. Parameters $\zeta^{\text{cool}}_\kappa$ and $\zeta^{\text{heat}}_\kappa$ are positive material constants for the cooling and heating cycle, respectively. The volume fraction $\chi^0_{\kappa,C}$ is introduced to ensure the cyclic description and the continuity of the laws. It is equal to the value of the $\kappa$-crystalline volume fraction at reversal points – when a passage from heating to cooling (or vice-versa) takes place – and is chosen to be given as (Arricca et al., 2024)

$$\chi^0_{\kappa,C} = \begin{cases} \chi^{max}_{\kappa,C} - \chi^{\text{heat}}_{\kappa,C}, & \dot{T} \leq 0, \\ \chi^{\text{cool}}_{\kappa,C}, & \dot{T} > 0, \end{cases} \tag{55}$$

with $\chi^{\text{heat}}_{\kappa,C}$ and $\chi^{\text{cool}}_{\kappa,C}$ representing the volume fractions of crystalline phase available from previous heating and cooling cycle, respectively. For initialization purposes we state that

$$\chi_{\kappa,C}(t=0) = \frac{\chi^{max}_{\kappa,C}}{1 + \exp\left[\zeta^{\text{cool}}_\kappa (T - T_{c,\kappa})\right]}. \tag{56}$$

Temperatures $T^{\text{eff}}_{c,\kappa}$ and $T^{\text{eff}}_{m,\kappa}$ are the effective crystallization and melting temperatures of PCL and PEG. Their introduction allows us to account for different phenomena that can affect the crystallization process, *i.e.*, the application of an external stress in two-way shape-memory tests, the stress associated with the actuation process and with the (un)melted phase during stress-free shape-memory tests, and the kinetic delay induced by an increase of the cooling(heating) rate (Scalet et al., 2018; Inverardi et al., 2022). They are defined empirically as (Arricca et al., 2024)

$$T_{c,\kappa}^{eff} = T_{c,\kappa} + \tau_{c,\kappa} v_\kappa^{cool} + \omega_{\kappa,C} |\sigma^{end}| + \delta_{\kappa,C} |\sigma_{act}^{end}| + \varphi_{\kappa,C}(\chi_\kappa - \chi_{\kappa,C}^{heat}),$$  (57a)

$$T_{m,\kappa}^{eff} = T_{m,\kappa} + \tau_{m,\kappa} v_\kappa^{heat}.$$  (57b)

Parameters $\tau_{c,\kappa} < 0$ and $\tau_{m,\kappa} > 0$ describe the linear shift in the transition temperatures (Scalet et al., 2018), which depend on the cooling and heating rate, $v_\kappa^{cool}$ and $v_\kappa^{heat}$, respectively. $\omega_{\kappa,C} > 0$ accounts for the stress influence on the crystallization process, with $\sigma^{end}$ value of the stress at the end of the isothermal loading, whereas $\delta_{\kappa,C} > 0$ describes the shift of transformation temperature due to stress (49), with $\sigma_{act}^{end}$ stress generated at the end of the previous heating cycle. The introduction of $\varphi_{\kappa,C}$ establishes the dependence of $T_{c,\kappa}^{eff}$ on the $\kappa$-crystalline phase.

The equations to describe the evolutions of the inelastic strain (46) are split as

$$\dot{\varepsilon}_C^{in} = \dot{\varepsilon}_{PCL,C}^{in} + \dot{\varepsilon}_{PEG,C}^{in}, \qquad \dot{\varepsilon}^{st,f} = \dot{\varepsilon}_{PCL}^{st,f} + \dot{\varepsilon}_{PEG}^{st,f}.$$  (58)

Here, $\dot{\varepsilon}_C^{in}$ denotes the inelastic strain evolution induced by the formation of PCL and PEG crystals, which generate strain ratios $\dot{\varepsilon}_{PCL,C}^{in}$ and $\dot{\varepsilon}_{PEG,C}^{in}$. Rather, $\dot{\varepsilon}_C^{st,f}$ describes the deformation ratio induced by the high-temperature loading, temporarily stored at low temperatures, in the individual PCL and PEG crystal domains, via $\dot{\varepsilon}_{PCL}^{st,f}$ and $\dot{\varepsilon}_{PEG}^{st,f}$.

Strain $\varepsilon_{\kappa,C}^{in}$ evolves in time according to the following law (Scalet et al., 2018)

$$\dot{\varepsilon}_{\kappa,C}^{in} = \begin{cases} \dot{\chi}_{\kappa,C}\left(\dfrac{\sigma}{\alpha_\kappa} + \gamma_\kappa \sigma_{act}^{end} + \theta_\kappa \dfrac{\sigma_{act}^{end}}{|\sigma_{act}^{end}|}\right), & \dot{T} < 0, \\[1em] \dfrac{\dot{\chi}_{\kappa,C}}{\chi_{\kappa,C}} \varepsilon_{\kappa,C}^{in}, & \dot{T} > 0, \\[1em] 0, & \dot{T} = 0, \end{cases}$$  (59)

where terms $\sigma \alpha_\kappa^{-1}$ and $\gamma_\kappa \sigma_{act}^{end}$ account for the contribution of the applied stress and the stress generated at the end of the previous heating cycle, respectively. Term

$$\theta_\kappa = \theta_{\kappa,C} \chi_{\kappa,C}^{heat} + \theta_{\kappa,A}(\chi_\kappa - \chi_{\kappa,C}^{heat})$$  (60)

allows the consideration of the effect of the $\kappa$ amorphous and crystalline phases by means of parameters $\theta_{\kappa,A}$ and $\theta_{\kappa,C}$, respectively.

Strain $\varepsilon_\kappa^{st,f}$ stores the deformation induced by the applied load, varying under isothermal loading, for $T > T_{m,\kappa}$. It is expected to decrease during heating under stress-free conditions, and kept constant during heating under an applied stress. Accordingly, the following evolution laws are chosen (Arricca et al., 2024)

$$\dot{\varepsilon}_\kappa^{st,f} = \begin{cases} \beta_\kappa \dot{\varepsilon}_{\kappa,A}^{el}, & \dot{T} = 0 \text{ and } T > T_{m,\kappa}, \\[1em] \dfrac{\dot{\chi}_{\kappa,C}}{\chi_{\kappa,C}} \varepsilon_\kappa^{st,f}, & \dot{T} > 0 \text{ and } |\sigma| = 0, \\[1em] 0, & \text{otherwise}, \end{cases}$$  (61)

with $\beta_\kappa$ material parameter and $\dot{\chi}_{\kappa,C}$ provided by Eqs. (54).

The numerical treatment of model Eqs. (54), (59) and (61) is performed by employing a backward-Euler integration algorithm and the solution of the involved nonlinear systems are obtained by means of the function *fsolve* implemented in the MATLAB environment.

### 5.5. Model calibration

The proposed model presents the following material parameters ($\kappa$ = PCL, PEG): *(i)* two elastic moduli for the amorphous and crystalline phases of the entire network, $E'_A T$ and $E_C$; *(ii)* the crystallization and melting temperatures, $T_{c,\kappa}$ and $T_{m,\kappa}$ (Table 1); *(iii)* the crystalline phase evolution parameters represented by the mass densities of the amorphous and crystalline domains, $\varrho_{\kappa,A}$ and $\varrho_{\kappa,C}$ (Section 5.1), two maximum crystallinity weight contents, $\chi_{\kappa,C}^{w,max}$ (Table 1), two maximum crystallinity volumetric fractions, $\chi_{\kappa,C}^{max}$ (evaluated via Eq. (36)), two parameters $\zeta_\kappa^{cool}$, and two parameters $\zeta_\kappa^{heat}$; *(iv)* the thermal expansion coefficients of the entire network, $\alpha_A$ and $\alpha_C$, and the thermal strain increment, $\Delta \varepsilon_C^{th}$; *(v)* two heating rate parameters, $\tau_{m,\kappa}$, and two cooling rate parameters $\tau_{c,\kappa}$; *(vi)* the evolution parameters $\alpha_\kappa$, $\gamma_\kappa$, $\theta_{\kappa,A}$ and $\theta_{\kappa,C}$, and $\beta_\kappa$; *(vii)* the temperature parameters $\omega_{\kappa,C}$, $\delta_{\kappa,C}$, and $\varphi_{\kappa,C}$.

The mechanical parameters related to the hyperelastic responses of the amorphous and crystalline phases, $E'_A$ and $E_C$, have been derived from tensile tests above the melting and below the crystallization temperature, $T_m = T_{m,PCL}$ and $T_c = T_{c,PEG}$, according to experimental evidences — see Eq. (34) and Table 1. Fig. 6 shows the identification process of $E'_A$ for the different weight ratios, 1:2, 1:1, and 2:1. Evaluations are performed on the experimental curves (Inverardi et al., 2022) visible in Fig. 6(a), and individual calibrations of $E'_A$ for the three different weight ratios analyzed are shown in Figs. 6(b)–6(d). The calibration of $E_C$ is rather carried out in such a way that the maximum total modulus (47) at $T = -20\,^\circ\mathrm{C}$ coincides with experimental data (Inverardi et al., 2022).

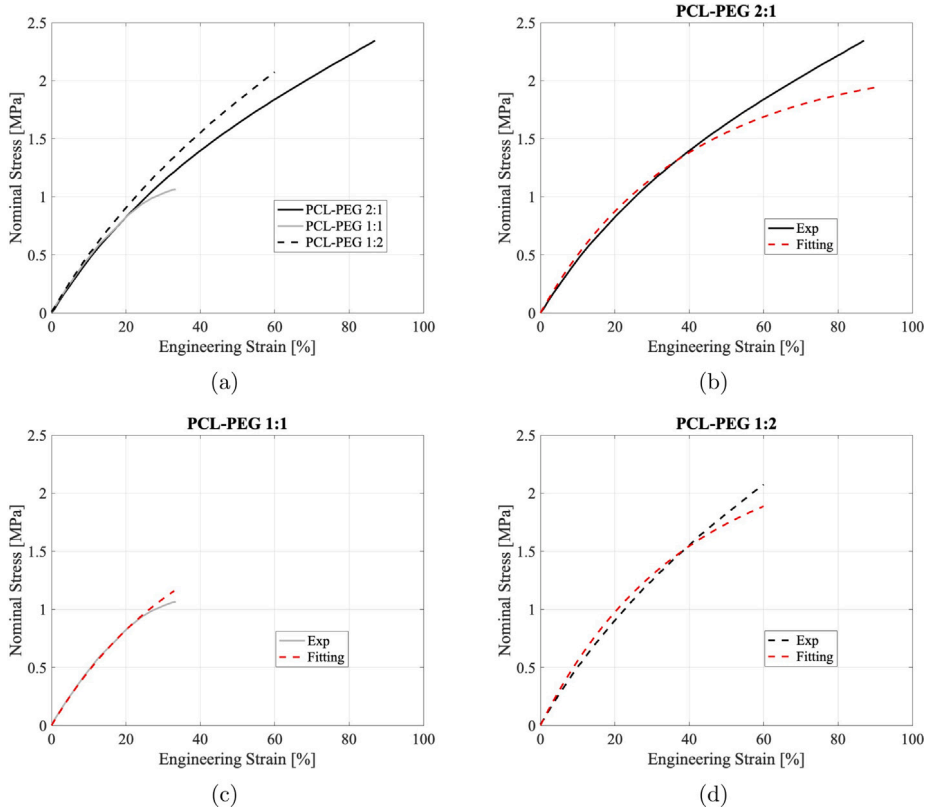

**Fig. 6.** (a) Experimental curves reproduced from Inverardi et al. (2022) and evaluation of $E'_A$ for the PCL-PEG weight ratio (b) 2:1, (c) 1:1 and (d) 1:2.

The crystalline phase evolution parameters for the cooling and heating cycles, $\zeta^{\mathrm{cool}}_{\mathrm{PCL}}$ and $\zeta^{\mathrm{cool}}_{\mathrm{PEG}}$, $\zeta^{\mathrm{heat}}_{\mathrm{PCL}}$ and $\zeta^{\mathrm{heat}}_{\mathrm{PEG}}$, are calibrated by means of model validation on the crystalline content-temperature curves shown in Fig. 7, and are listed in Table 3. Here, we have evaluated $\chi^w_C = \chi^w_{\mathrm{PCL},C} + \chi^w_{\mathrm{PEG},C}$ making use of Eq. (37), evolution laws (54), and Eqs. (55) and (56) after identification of the maximum PCL and PEG crystalline volume contents via Eq. (36).

The thermal expansion coefficients, $\alpha_A$ and $\alpha_C$, and the thermal strain increment, $\Delta\varepsilon^{th}_C$, for the three weight ratios analyzed, are calibrated via the model validation on the engineering strain–temperature curves for the two-way SME under a constant applied stress shown in Fig. 8. The aforementioned mechanical and thermal parameters, which are related to the whole PCL-PEG network, are listed in Table 2.

Table 3 shows all the remaining parameters calibrated in the model. Heating rate parameters $\tau_{\mathrm{m,PCL}}$ and $\tau_{\mathrm{m,PEG}}$, cooling rate parameters $\tau_{\mathrm{c,PCL}}$ and $\tau_{\mathrm{c,PEG}}$, evolution parameters $\alpha_{\mathrm{PCL}}$, $\alpha_{\mathrm{PEG}}$, $\beta_{\mathrm{PCL}}$, $\beta_{\mathrm{PEG}}$, $\theta_{\mathrm{PCL},A}$, $\theta_{\mathrm{PCL},C}$, $\theta_{\mathrm{PEG},A}$ and $\theta_{\mathrm{PEG},C}$, and temperature parameters $\omega_{\mathrm{PCL},C}$ and $\omega_{\mathrm{PEG},C}$, are calibrated on the two-way curve illustrated in Fig. 8. Evolution parameters $\gamma_{\mathrm{PCL}}$ and $\gamma_{\mathrm{PEG}}$, and effective crystallization temperature parameters $\delta_{\mathrm{PCL},C}$ and $\delta_{\mathrm{PEG},C}$, $\varphi_{\mathrm{PCL},C}$ and $\varphi_{\mathrm{PEG},C}$, are calibrated balancing the numerical results obtained for the engineering strain–temperature curves shown in Fig. 9, which demonstrate the reversibility of the two-way SME, and results obtained for the stress-free two-way SME, illustrated in Fig. 10.

### 5.6. Two-way shape-memory tests

We here show the validation of the model illustrating the two-way shape-memory behavior of the thermal-responsive PCL-PEG network under applied stress and stress-free conditions. Interested readers can find the detailed protocols adopted in experimental tests in Inverardi et al. (2022).

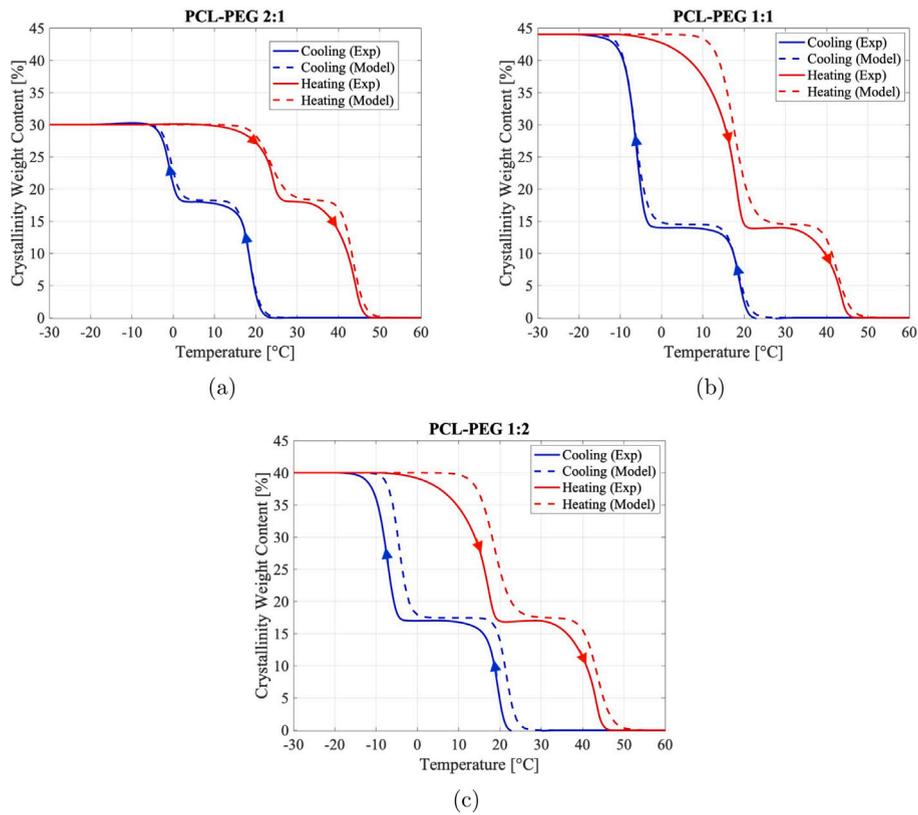

**Fig. 7.** Crystalline weight content vs. temperature curves for the PCL-PEG weight ratio (a) 2:1, (b) 1:1 and (c) 1:2.

**Table 2**
Model parameters of the PCL-PEG network adopted in the numerical simulations.

| Parameter | PCL-PEG 2:1 | PCL-PEG 1:1 | PCL-PEG 1:2 | Units |
|---|---|---|---|---|
| $E'_A$ | 0.017 | 0.016 | 0.019 | MPa K$^{-1}$ |
| $E_C$ | 618.3 | 879.7 | 1062.4 | MPa |
| $\alpha_A$ | $0.53 \cdot 10^{-4}$ | $1.3 \cdot 10^{-4}$ | $2.1 \cdot 10^{-4}$ | °C$^{-1}$ |
| $\alpha_C$ | $6.5 \cdot 10^{-4}$ | $4 \cdot 10^{-4}$ | $3.35 \cdot 10^{-4}$ | °C$^{-1}$ |
| $\Delta\varepsilon_C^{th}$ | $5.5 \cdot 10^{-3}$ | $5.8 \cdot 10^{-4}$ | $6.5 \cdot 10^{-3}$ | – |

### 5.6.1. Two-way shape-memory behavior under applied stress

For the two-way shape-memory behavior under an applied stress, the specimen is first kept at temperature $T = 65\,°\text{C}(>T_\text{m})$, and the deformation is carried out by means of the application of a certain value of engineering (nominal) stress to provide a deformation of 20%. By keeping the applied stress constant, a shape-memory cycle is then performed under the applied stress by first cooling the specimen until $T = -20\,°\text{C}(<T_\text{c})$ at 2 °C min$^{-1}$ and then performing a heating process until the initial temperature. The evolution of the engineering strain, $\bar{\varepsilon} = \lambda - 1$, as a function of the variation of temperature, is shown in Fig. 8 for the three weight ratio combinations of PCL and PEG analyzed.

Prediction capabilities of the model for such cases are of undeniable acceptance. In particular, the model is able to well describe the sequential elongations and contractions during cooling and heating, respectively. Upon cooling, a moderate elongation first occurs before the crystallization takes place; then, the so-called CIEs begin when passing the crystallization temperature of PCL, first, and of PEG, after. The inflection points of the curves defining the CIEs for the PCL and the PEG, well catched by the model, are known as effective crystallization temperatures and are slightly higher than the crystallization temperature, $T_\text{c,PCL}$ and $T_\text{c,PEG}$, measured through DSC tests, due to the presence of the applied stress (Inverardi et al., 2022). Upon heating, thermal contraction first occurs, followed by the so-called MICs when passing the melting temperature of PEG, first, and of PCL, after. At the end of the heating step, a complete recovery of the strain is evident. The inflection points of the curves defining the MICs for PCL and PEG are also well described by the model and are known as effective melting temperatures. They are slightly higher than the melting temperature, $T_\text{m,PCL}$ and $T_\text{m,PEG}$, measured through DSC tests, due to the effect of the heating rate (Inverardi et al., 2022).

**Table 3**
Identified model parameters of the PCL-PEG system under investigation.

| Parameter | PCL-PEG 2:1 | PCL-PEG 1:1 | PCL-PEG 1:2 | Units |
|---|---|---|---|---|
| $\zeta_{PCL}^{cool}$ | 0.8 | 0.75 | 0.9 | °C$^{-1}$ |
| $\zeta_{PCL}^{heat}$ | 0.75 | 0.65 | 0.6 | °C$^{-1}$ |
| $\zeta_{PEG}^{cool}$ | 0.95 | 0.7 | 0.85 | °C$^{-1}$ |
| $\zeta_{PEG}^{heat}$ | 0.5 | 0.5 | 0.5 | °C$^{-1}$ |
| $\tau_{c,PCL}$ | −2.3 | −1 | −4.7 | min |
| $\tau_{m,PCL}$ | 4.5 | 5.3 | 3.9 | min |
| $\tau_{c,PEG}$ | −3.5 | −0.5 | −4 | min |
| $\tau_{m,PEG}$ | 9.3 | 10.8 | 8.9 | min |
| $\alpha_{PCL}$ | 2 | 1.54 | 4.5 | MPa |
| $\beta_{PCL}$ | 0.63 | 0.63 | 0.63 | – |
| $\gamma_{PCL}$ | $8 \cdot 10^{-3}$ | $6.5 \cdot 10^{-3}$ | $3.5 \cdot 10^{-3}$ | MPa$^{-1}$ |
| $\alpha_{PEG}$ | 8 | 13.2 | 4.4 | MPa |
| $\beta_{PEG}$ | 0.6 | 0.58 | 0.65 | – |
| $\gamma_{PEG}$ | $1.9 \cdot 10^{-3}$ | $0.1 \cdot 10^{-3}$ | $0.7 \cdot 10^{-3}$ | MPa$^{-1}$ |
| $\theta_{PCL,C}$ | 1.1 | 0.25 | 0.55 | – |
| $\theta_{PCL,A}$ | 0.09 | 0.02 | 0.05 | – |
| $\theta_{PEG,C}$ | 1 | 0.24 | 0.3 | – |
| $\theta_{PEG,A}$ | 0.05 | 0.04 | 0.01 | – |
| $\omega_{PCL,C}$ | 6.5 | 4.9 | 5.5 | °C MPa$^{-1}$ |
| $\delta_{PCL,C}$ | 0.1 | 0.15 | 0.16 | °C MPa$^{-1}$ |
| $\varphi_{PCL,C}$ | 6.5 | 5 | 5.5 | °C MPa$^{-1}$ |
| $\omega_{PEG,C}$ | 13.5 | 12.5 | 13.5 | °C MPa$^{-1}$ |
| $\delta_{PEG,C}$ | 0.06 | 0.07 | 0.07 | °C MPa$^{-1}$ |
| $\varphi_{PEG,C}$ | 12.5 | 4.5 | 7.5 | °C MPa$^{-1}$ |

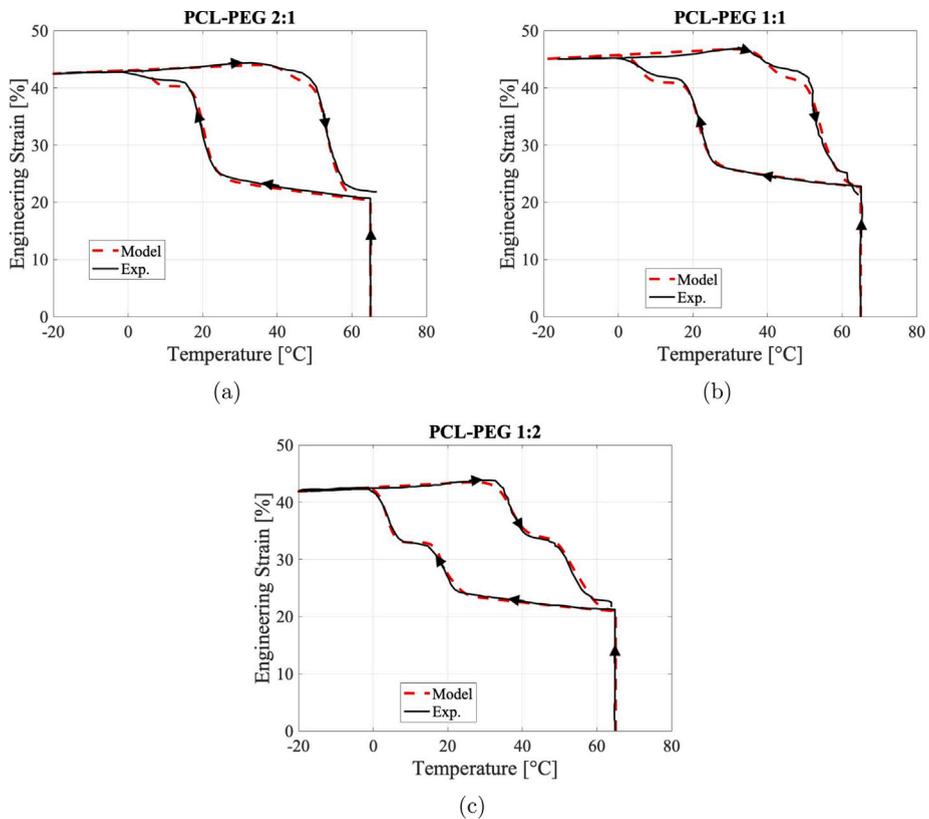

**Fig. 8.** Model validation on engineering strain vs. temperature curves for the PCL-PEG weight ratio (a) 2:1, (b) 1:1 and (c) 1:2 for the two-way SME under applied stress.
*Source:* Experimental curves are taken from Inverardi et al. (2022).

#### 5.6.2. Two-way shape-memory behavior under stress-free conditions

For the two-way shape-memory behavior under stress-free conditions, Figs. 9 and 10 show the variation of the engineering strain, $\bar{\varepsilon} = \lambda - 1$, as the temperature varies, for the three weight ratio combinations of PCL and PEG analyzed. The specimen is kept above the melting temperature, $T = 65\,°C\,(>T_m)$, and the application of a certain engineering stress generates a deformation of about 20%. Shape-memory cycles are performed via the application on the deformed specimen of a cooling step up to $T = -20\,°C\,(<T_c)$ at a velocity of $2\,°C\,min^{-1}$, while applied stress is kept constant. Here, stress is removed, and a first heating step up to a certain temperature $T_{act}$ is followed by subsequent cooling–heating cycles under stress-free conditions at a cooling and heating rates of $2\,°C\,min^{-1}$. The deformation is eventually fully recovered during the last heating process until the initial temperature, where the specimen restores its permanent shape.

Fig. 9 reports the comparison between numerical and experimental curves for tests applying $T_{act} = 43\,°C$, while Fig. 10 verifies model capabilities in the case of varying $T_{act}$. On the one hand, $T_{act}$ was assumed equal to $43\,°C$ in Fig. 9 to exploit the (fully melted) PEG phase as actuation domain which cyclically crystallizes and melts under the subsequent stress-free thermal cycles, inducing the change in strain. The (fully unmelted) PCL phase provides the internal stress for the subsequent two-way actuation through its skeleton-like domain. On the other hand, in Fig. 10 the different values of $T_{act}$ allow us to discuss model performances in correctly describing the effect of partially melted PEG and PCL phases on the achievable actuation strains under subsequent stress-free thermal cycles. In detail, for the highest $T_{act}$ the two-way mechanism involves the fully melted PEG phase and the melted part of the PCL phase to act as the actuation domain, whereas the unmelted PCL crystals provide the crystalline skeleton domain. The amount of the actuation strain thus depends on $T_{act}$, since it influences the contents of the melted/unmelted crystals.

The prediction capabilities of the model are reasonably acceptable, especially in view of the complexity of phenomena characterizing the two-way SME under stress-free conditions, as the actuation strain magnitude and crystallization process onset and termination (Arricca et al., 2024). Specifically, the model is able to describe the experimental amount of actuation strains under stress-free conditions for the different $T_{act}$. Also, it is possible to note that the temperatures of the CIEs in the reversible cycles without load are reasonably predicted for all the material systems. These inflection points are influenced by the generated internal stress and unmelted crystals.

It is further worth to underline the complex system behavior and material response variability, as well as the difficulties in the experimental setup and testing. In the latter regard, note the differences between numerical and experimental results for the weight ratios of 1:1 and 2:1 in Figs. 9(b) and 9(c), and, although less relevant, for the weight ratios of 1:1 and 2:1 in Figs. 10(b) and 10(c). Consider that model calibration of the strain–temperature behavior occurs by means of the previous case of the two-way shape-memory behavior under applied stress, therefore the first cooling-heating steps in Figs. 9 and 10 are coincident with results reported in Fig. 8. The main reason for the imperfect match between the model and experimental results may be reasonably attributed to the complex, articulated multi-step experimental setup and testing. Differences can be ascribed to a measurement error or to the fact that applied deformations are actually obtained by the application of stresses, and the values of stress adopted to obtain experimental results are not always the same. Another reason underlying such discrepancy may be attributed to the experimental difficulty in keeping constant the cooling-heating rates, for such complex thermo-mechanical histories, involving several changes between cooling and heating segments. Further slight discrepancy may derive from the numerical model, where parameters have been fitted with a trial-and-error procedure, but they could be optimized with an optimization procedure, which nonetheless falls beyond the scope of the current work.

### 6. Conclusions

This paper introduces a phenomenological framework, rooted in the finite strain theory of continuum mechanics and based on a phase transition approach, to describe the thermally-driven one-way and two-way shape-memory behavior of generic multi-phase semi-crystalline networks.

The description of the material system has allowed the selection of proper phase and control variables. Kinematic specifications and the constitutive theory have been established following rigorous and consolidated approaches in finite strain mechanics and elasto-plasticity. Thorough discussions on adopted procedures, restrictions arising from the Hencky and phase transition models, and proper insights on the free energy selection and stress definitions to characterize the stress–strain response of the material, have been provided. At the current state of the art, the complexity of the shape-memory behavior of these systems is far for being fully and univocally understood and modeled. We therefore believe that the present work, underlying the difficulties arising in SMP modeling by means of proper investigations and elucidations, opens new perspective for future insights and model developments to characterize the complex behavior of such innovative smart materials.

The present model is nonetheless accurate in the prediction of the behavior of these materials, as the validation of the framework against experimental results demonstrates. Results are indeed in agreement with experimental evidences for semi-crystalline networks exhibiting the one-way and the two-way SME under both stress and stress-free conditions, for strain ranges of about 40%–45% and temperature ranges between −20 and 65 °C.

Most interestingly, the model is completely general and can be easily adapted to various semi-crystalline networks and related polymers, by specifying the number of phases and by defining the free-energy functions and *ad-hoc* evolution laws based on experimental evidences. Accordingly, it furnishes a reliable tool for design purposes of structures and components based on these materials. Moreover, further developments may be focused on the extension of multi-phase material systems relying on the glass transition temperatures as switching mechanisms for the shape-memory behavior.

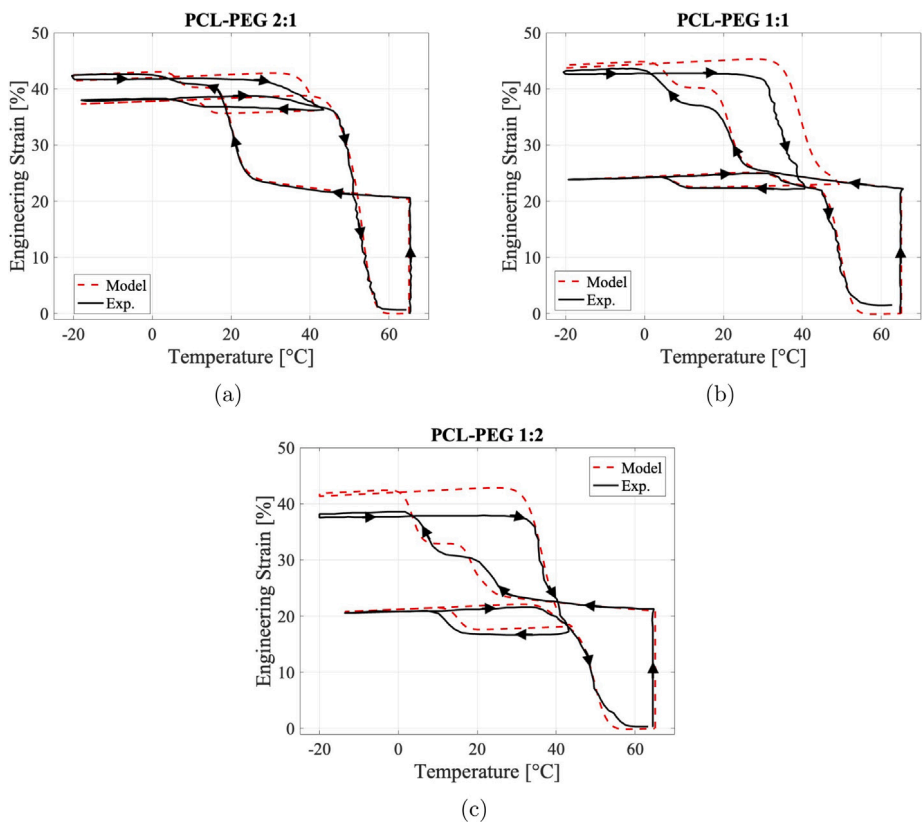

**Fig. 9.** Model validation on engineering strain vs. temperature curves for the PCL-PEG weight ratio (a) 2:1, (b) 1:1 and (c) 1:2., demonstrating the reversibility of the stress-free two-way SME.

**CRediT authorship contribution statement**

**Matteo Arricca:** Writing – review & editing, Writing – original draft, Visualization, Validation, Software, Methodology, Formal analysis, Data curation, Conceptualization. **Nicoletta Inverardi:** Writing – review & editing, Validation, Methodology, Investigation. **Stefano Pandini:** Writing – review & editing, Methodology, Investigation, Conceptualization. **Maurizio Toselli:** Writing – review & editing, Methodology, Investigation, Conceptualization. **Massimo Messori:** Writing – review & editing, Methodology, Investigation, Conceptualization. **Giulia Scalet:** Writing – review & editing, Writing – original draft, Supervision, Software, Project administration, Methodology, Funding acquisition, Data curation, Conceptualization.

**Declaration of competing interest**

The authors declare that they have no known competing financial interests or personal relationships that could have appeared to influence the work reported in this paper.

**Acknowledgments**

This work was funded by the European Union ERC CoDe4Bio Grant ID 101039467. Views and opinions expressed are however those of the author(s) only and do not necessarily reflect those of the European Union or the European Research Council. Neither the European Union nor the granting authority can be held responsible for them.

**Data availability**

Data generated in this study have been deposited in the Zenodo database at https://doi.org/10.5281/zenodo.14196360.

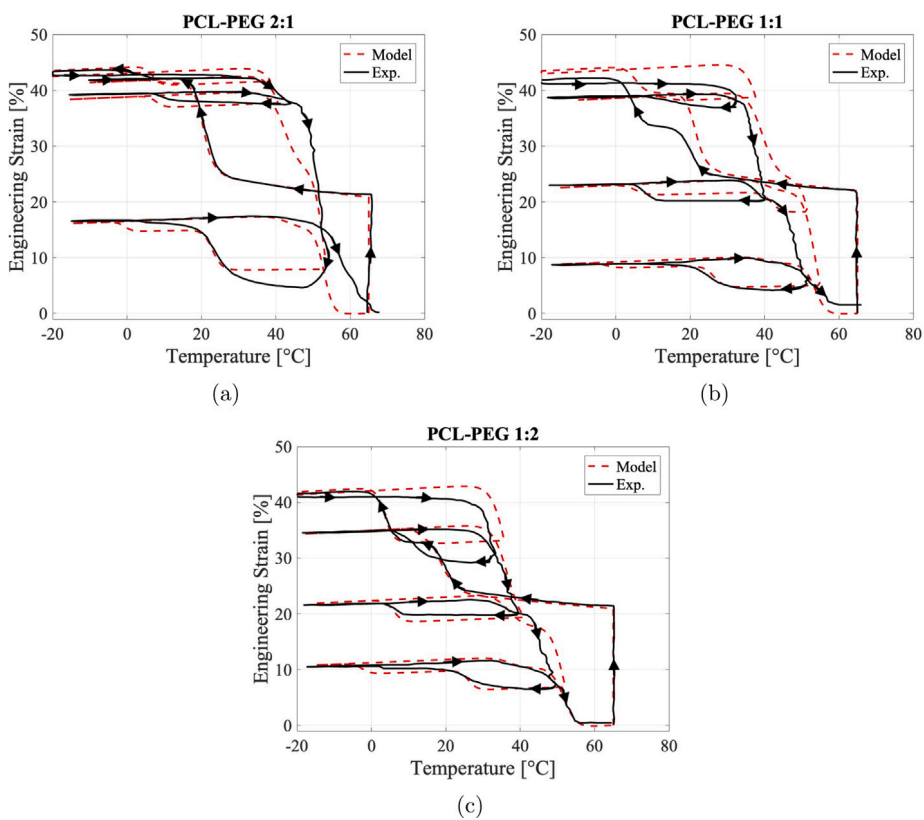

**Fig. 10.** Model validation on engineering strain vs. temperature curves for the PCL-PEG weight ratio (a) 2:1, (b) 1:1 and (c) 1:2., demonstrating the effect of the actuation temperature on the stress-free two-way SME.